\documentclass[%
  reprint,
superscriptaddress,
preprintnumbers,
 amsmath,amssymb,
 aps,
prb,twocolumn
]{revtex4-2}

\usepackage[normalem]{ulem}
\usepackage{amsmath}
\usepackage{bigints}
\usepackage{amsfonts}
\usepackage{multirow}
\usepackage{diagbox}
\usepackage{amssymb}
\usepackage{bm}
\usepackage{adjustbox}
\usepackage{epstopdf}
\usepackage{color}
\usepackage{xcolor}
\usepackage{textcomp}
\usepackage{cancel}
\usepackage{gensymb}
\usepackage{subfigure}
\usepackage{graphicx}
\usepackage{dcolumn}
\usepackage{bm}
\usepackage{hyperref}
\hypersetup{colorlinks=true,allcolors=blue}

\newcommand{\pmat}[1]{\begin{pmatrix}#1\end{pmatrix}}

\begin{document}
\title{Light-Driven Transitions in Quantum Paraelectrics}

\author{Zekun Zhuang}
 \email{zekun.zhuang@physics.rutgers.edu}
 \affiliation{Center for Materials Theory, Rutgers University,
Piscataway, New Jersey, 08854, USA}

 \author{Ahana Chakraborty}
 \email{ahana@physics.rutgers.edu}
 \affiliation{Center for Materials Theory, Rutgers University,
Piscataway, New Jersey, 08854, USA}

\author{Premala Chandra}
\email{pchandra@physics.rutgers.edu}
\affiliation{Center for Materials Theory, Rutgers University,
Piscataway, New Jersey, 08854, USA}
\author{Piers Coleman}
\email{coleman@physics.rutgers.edu}
\affiliation{Center for Materials Theory, Rutgers University,
Piscataway, New Jersey, 08854, USA}
\affiliation{Hubbard Theory Consortium, Department of Physics, Royal Holloway, University of London, Egham, Surrey TW20 0EX, UK}
\author{Pavel A. Volkov}
\email{pv184@physics.rutgers.edu}
\affiliation{Center for Materials Theory, Rutgers University,
Piscataway, New Jersey, 08854, USA}
\affiliation{Department of Physics, Harvard University, Cambridge, Massachusetts, 02138 USA}
\affiliation{Department of Physics, University of Connecticut, Storrs, Connecticut 06269, USA}

\date{\today }

\begin{abstract}
Motivated by recent experiments on pump-induced polar ordering in the quantum paraelectric SrTiO$_3$, we study a driven phonon system close to a second order phase transition. Analyzing its classical dynamics, we find that sufficiently strong driving leads to transitions into polar phases whose structures, determined by
the light polarization, are not all accessible in equilibrium.  In addition, for certain intensity
profiles we demonstrate the possibility of two-step transitions as a function of fluence. For even stronger field 
intensities, the possibility of period-doubling and chaotic behavior is demonstrated.  Finally we develop a generalized
formalism that allows us to consider quantum corrections to the classical dynamics in a systematic fashion. We predict a shift in the critical pump fluence due to quantum fluctuations with a characteristic dependence on the fluence increase rate
that should be observable in experiment.

\end{abstract}

\maketitle


\section{Introduction}

The control and design of properties in quantum materials are outstanding goals both to address fundamental questions and to develop applications with quantum advantages. Because the potential and the kinetic energy scales in these materials are
comparable, their quantum phases are very sensitive to external fields \cite{basov2017towards,Buzzi18,Bloch22}. Advances in the production of strong light pulses in mid-infrared and terahertz 
ranges \cite{hoffmann2011intense,kampfrath2013resonant,nicoletti2016nonlinear} have led to opportunities for 
 such light to strongly modify the low-energy physics of materials. In particular, 
light-induced electronic \cite{rini2007control,mankowsky2016non} and 
lattice \cite{forst2011nonlinear,nelson2019,cavalleri2019,Disa21,Abreu22,Henstridge22,Porer2018,Zhang2023} phase transitions \cite{zhang2014keynote,de2021colloquium,bao2022light} have been observed. 
 
Recently terahertz (THz) field-induced ferroelectricity has been demonstrated in SrTiO$_3$ (STO) \cite{nelson2019,cavalleri2019}, in agreement with semiclassical predictions based on nonlinear phonon coupling \cite{subedi2014,Subedi2017,itin2018}. Though this material remains paraelectric to the lowest temperatures \cite{muller1979}, its polar mode can be softened by chemical
substitution \cite{bednorz1984,itoh1999} and strain \cite{haeni2004room} leading to a polar instability.  
However, unlike these material modifications, the pump-induced phase transition occurs as a function of fluence.
Since quantum criticality is observed in $^{18}O$ doped STO \cite{Kvyatkovskii01,Rowley14,Chandra17}, 
there is also the intriguing possibility of driving non-equilibrium quantum critical dynamics in this quantum paraelectric.

For driven classical phase transitions, the creation of topological defects with universal scaling of their density has been predicted and observed in materials \cite{griffin2012scaling,ulm2013observation,lin2014topological}. Universal dynamics \cite{gagel2014,Gagel15,mitra2015} have emerged from theoretical studies of dynamical quantum critical effects, as have signatures of
dynamical quantum phase transitions such as the Loschmidt echo \cite{Jurcevic17,Heyl_2018,Heyl_2019}.  
However these characterizations have predominantly been  realized in closed quantum systems like cold atoms where
initial states can be carefully prepared \cite{Jurcevic17,Heyl_2018, Heyl_2019}. By definition, quantum materials are not isolated from their environments and their constituents, unlike those of their synthetic quantum counterparts,
cannot be easily addressed microscopically.

The light-induced ferroelectricity experiments \cite{nelson2019,cavalleri2019,Abreu22,Henstridge22} thus demand new ways
to model strong classical drive protocols that induce critical dynamics, both classical and quantum,
and to identify macroscopic signatures of dynamical quantum phase transitions.  
Theoretical studies suggest that many Thz field-induced phenomena may be due to nonlinear 
phonon interactions \cite{subedi2014,Subedi2017,itin2018,Knap16,Kennes17,Cantaluppi18,Shin20,Guan21,Grandi2021,Klein2020,Puviani2018}.
Recently many of the observed features
in the field-induced ferroelectricity experiments \cite{nelson2019,cavalleri2019} have been simulated  \cite{rubio2022}
with a time-dependent density functional theory analysis where the anharmonic coupling between the driven and the critical phonons is modelled by a Schr\"odinger-Langevin approach \cite{Shin21}. 
In parallel a Matsubara action analysis has been developed to describe an off-resonant drive-induced feroelectric transition \cite{balatsky2022}, where results have been obtained using a saddle-point (classical) calculation.

\begin{figure*}
\includegraphics[width=0.99\textwidth]{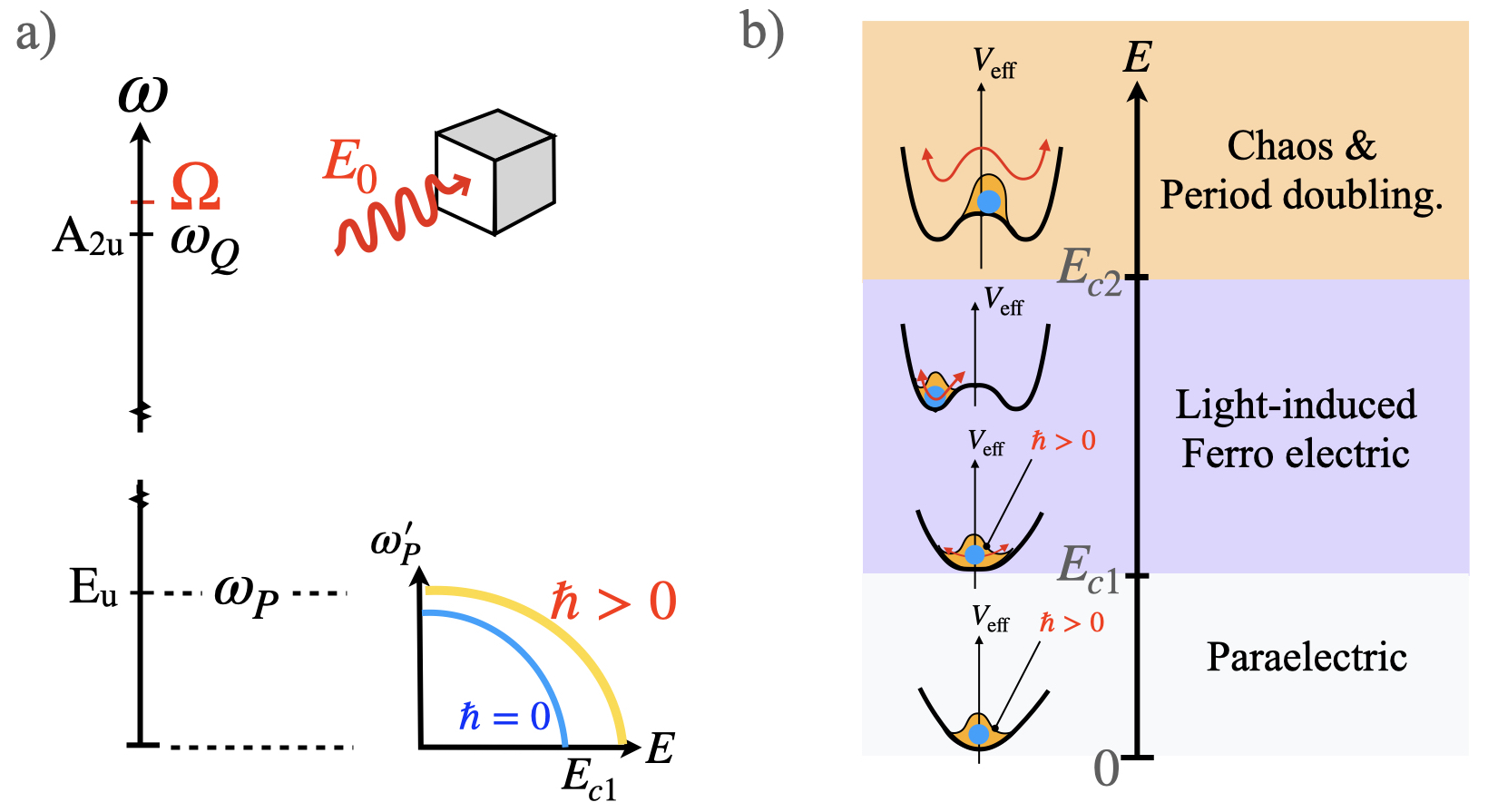}
  \caption{a) Schematic of the energy scales in a light-induced
  experiment, showing the pumping frequency $\Omega$ that resonantly drives
  the high frequency anharmonic A$_{2u}$ optical mode at frequency
  $\omega_Q$ and the low frequency soft polar E$_{u}$ mode. Shown at
  the bottom is the evolution of the soft polar mode frequency with
  fluence, and the effect of quantum fluctuations ($\hbar  >0 $) b) Phase
  transitions as a function of electric field amplitude, showing
  the evolution of an effective potential $V_{\rm eff}$ with increasing
  fluence. The orange wavepacket around the blue classical configuration
  represents the effect of quantum fluctuations.  ($\hbar > 0 $)  } 
   \label{introfig}
   \end{figure*}

The key idea of light-induced phase transitions is a  generalization of optical tweezers \cite{Ashkin70,Ashkin80,Bustamante21} to many-body physics \cite{basov2017towards,Buzzi18}.  In the context of optical tweezers, a high frequency laser mode polarizes the atoms, reducing their energy by an amount proportional to the intensity of the light, producing an effective potential 
$
V_{\text{eff}}({\bf x}) = - \frac{1}{2}\chi_0 E({\bf x})^2,
$
where $\chi_0$ is the polarizability of the atom. 
Similarly, in light-induced phase transitions, the intensity of a high frequency laser modifies the effective potential of a polar soft mode to be 
\begin{equation}
V_{\text{eff}}(P)= \frac{1}{2}(\omega_P^2- \chi E^2)P^2+ \frac{u}{2}P^4,
\label{EffV}
\end{equation}
where $P$ and $\omega_P$ are the the polarization and frequency of the soft mode respectively, $u$ is the quartic coefficient and $\chi$ is the coupling to the electric field intensity $E^2$. Once the shifted soft mode frequency 
\begin{equation}
\omega_P^2(E) = \omega_P^2 - \chi E^2 
\label{omegaE}
\end{equation}
vanishes, a phase transition into a broken symmetry state with finite polarization magnitude
\begin{equation}
|P_0| = \sqrt{\frac{-\omega^2_P(E)}{2 u}}
\label{P0}
\end{equation} 
occurs.  

Here we consider a harmonic driving electric field $E(t) = E_0 \cos \Omega t$.
\color{black} Typically, the coupling $\chi$ in \eqref{EffV} and \eqref{omegaE} is enhanced by resonantly driving an intermediate high frequency optical phonon that is anharmonically coupled to the polar mode (see Figure \ref{introfig} (a)). This process modifies the effective potential of the soft mode, ultimately inducing phase transitions as a function of fluence (see Figure \ref{introfig}b).  More specifically,
in addition to the ferroelectric transition, at higher field intensities the polarization 
fluctuations, $\delta P = |P - P_0|$,  become sufficiently large that the system oscillates between the two potential wells ($\delta P >> |P_0|$) and the system returns to being paraelectric on average. Qualitatively this is because at high
fields $P_0$ grows linearly with $E$ whereas $\delta P$ increases superlinearly. The latter occurs due to a
field-induced hardening of the soft polar mode frequency, that brings it closer to the pump frequency, enhancing the oscillation amplitude. This behavior results in a critical field $E_{c2}$ 
where $\delta P \approx P_0$ (see Figure \ref{introfig}b).

Since quantum criticality has been observed in a number of quantum paraelectrics at low tempertures \cite{Kvyatkovskii01,Rowley14,Chandra17},
it is natural to explore how the presence of quantum fluctuations will modify the critical fluence $E_{c1}$ 
(see Figure \ref{introfig}b) into the polar phase. 
Qualitatively we expect
the renormalized mass $\tilde{m} (t)$, the renormalized quadratic coefficient in the effective potential \eqref{EffV}, 
to have the form
\begin{equation}
\tilde{m} (t) = \omega_P^2 + m_{Cl}(t)  + m_Q (t)
\label{remass}
\end{equation}
where the time-dependence of the classical mass corrections, $m_{Cl} (t) = -\chi \{E(t)\}^2$ (cf. \eqref{omegaE}), results from the harmonic drive; $m_Q(t)$ refers to the time-dependent quantum mass corrections.
Since quantum fluctuations are expected to disorder the system, $m_{Cl}(t)$ and $m_Q(t)$ 
in \eqref{remass} act in opposition leading to a shift in the critical point. 
\color{black}
Nonequlibrium quantum dynamics has been previously studied 
after a quench to the quantum critical point \cite{gagel2014,Gagel15,mitra2015}, but here a new approach
is required to treat  dynamical quantum fluctuations when the classical order parameter is finite.  

%
 
In this paper, we present a theoretical study of light-induced transitions in quantum paraelectrics
where we explore their classical dynamics \cite{subedi2014,Subedi2017,itin2018} with controlled quantum corrections.
More specifically, classically we consider the effects of light polarization, long-range Coulomb interactions and drive fluence for the case of a resonantly driven phonon coupled to the soft mode (Sec. \ref{sec:model}). For a paraelectric with cubic symmetry (Sec. \ref{SecClassical}), we demonstrate fluence-and polarization dependent transitions into different ordered phases, some inaccessible in equilibrium:  for example we find two successive transitions as a function of fluence can occur in contrast to the one-stage symmetry-breaking routinely observed.
Finally, we demonstrate (Sec. \ref{sec:quant}) that the classical dynamical equations \cite{subedi2014,Subedi2017,itin2018}
emerge naturally within a Keldysh field theory. The quantum corrections to these equations can be then be treated systematically in a diagrammatic expansion.  We find that the critical fluence to enter the ferroelectric phase $E_{c1}$ 
(see Figure \ref{introfig}b) is shifted due to quantum fluctuations, a prediction that should be accessible in experiment.
 Here we have described light-induced transitions as a function of field intensity; an increase
in fluence leads to a decrease of $m_{Cl}(t)$ such that $\tilde{m}(t)$ will change sign as a function of time.  Thus driven transitions as a function of fluence and of time are closely connected,
and this link will be pursued particularly in our study of quantum effects.

\color{black}


\section{The Classical Action with Cubic Symmetry}
\label{sec:model}

We consider a three-dimensional paraelectric system with cubic symmetry group $O_h$, where the dipole moment corresponds to the three-fold degenerate irreducible representation $t_{1u}$. There are a number of cubic quantum paraelectrics including
KTaO$_3$, and we note that STO has a weakly distorted tetragonal structure. We assume that the only relevant phonon modes are two sets of $t_{1u}$ optical phonon modes, $P_i$ and $Q_i$ ($i=x,y,z$), where the $P$ and $Q$  are soft polar and
 higher energy modes respectively.  Consideration of two sets of modes is both motivated by experiment \cite{cavalleri2019}, and, as shown below, is necessary to provide a finite lifetime to the transient ferroelectric state after the pump is turned off, consistent with observation \cite{nelson2019}. The general classical action for $P_i(x,t)$ and $Q_i(x,t)$ has the  form
\begin{equation}
    S=S_2+S_4+S_{C}+S_{PQ}+S_E. \label{ModelS}
\end{equation}
Here  $S_2$ describes the harmonic terms in the action
\begin{align}
    S_2=\int_{x,t} &\sum_i\frac{1}{2}\left[(\partial_t Q_i)^2-c_q^2(\nabla Q_i)^2-\omega_Q^2 Q_i^2\right.\\\nonumber
    &+\left.(\partial_t P_i)^2-c_p^2(\nabla P_i)^2-\omega_P^2P_i^2 \right],
\end{align}
where $\omega_Q\gg \omega_P$ are the frequencies of the two optical modes, and $c_p$ and $c_q$ are their sound velocities 
respectively.  We will also use the shorthand notation $\int_{x,t}\equiv \int d^3x\int dt$ hereafter. $S_4$ represents the anharmonic interactions of the modes, taken to be local:
\begin{align}\label{quartic}
    S_4=-\int_{x,t} &\left[u_q(\sum_i Q_i^2)^2+v_q\sum_i Q_i^4 \right.\\\nonumber
    &+\left.u_p(\sum_i P_i^2)^2+v_p\sum_i P_i^4\right], 
\end{align}
where we require $v_{p(q)}>-u_{p(q)}$ if $u_{p(q)}>0$ and $v_{p(q)}>-3u_{p(q)}$ if $u_{p(q)}<0$ so that the energy is bounded from below.

The term  $S_C$  describes the Coulomb interaction between the charge fluctuations induced by the longitudinal fluctuations of the $P$ and $Q$ modes, which in reciprocal space is given by 

\begin{equation}
    \begin{gathered}
        S_{C}=-2 \pi\int_{k,t}  \frac{\rho(k) \rho(-k)}{k^2}
    \\
    \rho(k) =i\sum_{i} Z_P k_i P_i(k)+Z_Q k_i Q_i(k)
    \end{gathered}
    \label{CoulombS}
\end{equation}
where $Z_{Q(P)}$ is proportional to the effective charge of the $Q(P)$ mode, and we denote 
$\int _{k,t} = \int \frac{d^3k}{(2 \pi)^3}dt$. This term is responsible for the splitting between the longitudinal and transverse optical modes (LO-TO splitting).

$S_{PQ}$, the third term in Eq. (\ref{ModelS}), describes the nonlinear interaction between the $P$ and $Q$ modes. Only coupling between even powers of $P$ and $Q$ leads to qualitatively new effects; by contrast, cubic-linear or linear-linear couplings can be shown to simply renormalize the effects of linear coupling to the electric field (see Appendix \ref{app:PQeffects} for more discussions). In particular, these terms will lead to contributions $\propto Q(t), Q^3(t), P^2(t) Q(t)$ in the equation of motion for the $P$ mode. Since $Q$ mode is the one being driven by light, these terms will oscillate at the driving frequency (and its multiples) with a zero average. As we show below, such terms are not important for the determination of the onset of the pump-induced ferroelectricity and therefore can be neglected (see also discussion in Sec. \ref{TCOM}).
Therefore, we restrict ourselves to couplings with even powers of $P$ and $Q$, consistent with cubic symmetry:
\begin{equation}
    \label{PQcoupling}
  S_{PQ}= \int_{x,\tau}\left[\frac{\gamma_1}{2}|\Vec{P}|^2|\Vec{Q}|^2+\frac{\gamma_2}{2}(\Vec{P}\cdot\Vec{Q})^2+\frac{\gamma_3}{2}\sum_i P_i^2Q_i^2\right]\nonumber  
\end{equation}
    Finally, 
    \begin{equation}
        S_E = \int_{x,t} Z_q\vec{E}\cdot\vec{Q}+Z_p\vec{E}\cdot\vec{P}
        \label{eq:extfield}
    \end{equation}
    describes the interaction of a high frequency external driving field $E$ with the $P$ and $Q$ modes. Importantly, one observes that the light only couples to oscillations of $Q$ or $P$ with antiparallel wavevectors, i.e. $\vec{E}(\vec{k})$ couples to 
    $\vec{Q}(-\vec{k})$ and $\vec{P}(-\vec{k})$, and does not couple to oscillations orthogonal to $\vec{E}$. Since electromagnetic waves are transverse ($\vec{E}\cdot\vec{k}=0$), the external field couples  to transverse modes so that  $S_C$ \eqref{CoulombS} vanishes for the transverse optical modes relevant for our present study due to its longitudinal origin ($\rho(k) \propto \vec{k} \cdot \vec{E}$).

\section{Classical dynamics} \label{SecClassical}

In this Section we explore the equations of motion of the $P$ and the $Q$ modes that result from the classical
action with cubic symmetry just presented in Section \ref{sec:model}. We discuss the physically reasonable assumptions
we make so that these equations can be mapped onto that of a particle moving in an effective potential $V_\text{eff}$.
The instabilities of $V_\text{eff}$ are then studied as a function of light polarization (\ref{FEorder}). Intensity profiles
associated with second harmonic generation are next presented as experimental signatures of predicted polar phases often not
accessible in equilibrium (\ref{SHGsigs}).  

Finally (\ref{TCOM}) we go beyond the effective potential approximation and analyze
a minimalist model of two coupled scalar oscillators.  The resulting equation of motion is that of a generalized
Duffing oscillator and thus is expected to have rich dynamics \cite{Duffing18,strogatz2018,Chatterjee20}.  Indeed at large fluences, we find parameter regimes
where there are multiple steady state solutions and even chaotic behavior.  We also find persistance of the polar phase after the drive has ceased, in qualitative agreement with experiment \cite{nelson2019,cavalleri2019}.

\subsection{Effective Potential Approximation: Polarization-Controlled Ferroelectric Order}
\label{FEorder}

\begin{table*}[]
\begin{tabular}{cccccc}
\hline
\multicolumn{2}{|c|}{\multirow{2}{*}{}} & \multicolumn{2}{c|}{$2\tilde\gamma_1+\tilde\gamma_2+\tilde\gamma_3>0$} & \multicolumn{2}{c|}{$2\tilde\gamma_1+\tilde\gamma_2+\tilde\gamma_3<0$} \\ \cline{3-6} 
\multicolumn{2}{|c|}{} & \multicolumn{1}{c|}{$\tilde\gamma_1>0$} & \multicolumn{1}{c|}{$\tilde\gamma_1<0$} & \multicolumn{1}{c|}{$\tilde\gamma_1>0$} & \multicolumn{1}{c|}{$\tilde\gamma_1<0$} \\ \hline
\multicolumn{1}{|c|}{\multirow{2}{*}{$v_p>0$}} & \multicolumn{1}{c|}{$\tilde\gamma_2+\tilde\gamma_3>0$} & \multicolumn{1}{c|}{\begin{tabular}[c]{@{}c@{}}$[0 0 0]\rightarrow \{[1 1 0]\}_4$$^{a}$\\ $[0 0 0]\rightarrow\{[1 1 0]\}_4\rightarrow \{[a a c]\}_8$$^{b}$\end{tabular}} & \multicolumn{1}{c|}{\multirow{2}{*}{\begin{tabular}[c]{@{}c@{}}$[0 0 0]\rightarrow\{[1 1 0]\}_4$$^{a}$\\ $[0 0 0]\rightarrow\{[1 1 0]\}_4\rightarrow\{[a a c]\}_8$$^{b}$\end{tabular}}} & \multicolumn{1}{c|}{\multirow{4}{*}{\begin{tabular}[c]{@{}c@{}}$[0 0 0]\rightarrow\{[0 0 1]\}_2$$^{c}$\\ $[0 0 0]\rightarrow\{[0 0 1]\}_2\rightarrow\{[a a c]\}_8$$^{d}$\end{tabular}}} & \multicolumn{1}{c|}{\multirow{4}{*}{$[0 0 0]$}} \\ \cline{2-3}
\multicolumn{1}{|c|}{} & \multicolumn{1}{c|}{$\tilde\gamma_2+\tilde\gamma_3<0$} & \multicolumn{1}{c|}{$[0 0 0]\rightarrow\{[0 0 1]\}_2\rightarrow\{[a a c]\}_8$} & \multicolumn{1}{c|}{} & \multicolumn{1}{c|}{} & \multicolumn{1}{c|}{} \\ \cline{1-4}
\multicolumn{1}{|c|}{\multirow{2}{*}{$v_p<0$}} & \multicolumn{1}{c|}{$\tilde\gamma_2+\tilde\gamma_3>0$} & \multicolumn{1}{c|}{$[0 0 0]\rightarrow\{[100]\}_4$} & \multicolumn{1}{c|}{\multirow{2}{*}{$[0 0 0]\rightarrow\{[1 0 0]\}_4$}} & \multicolumn{1}{c|}{} & \multicolumn{1}{c|}{} \\ \cline{2-3}
\multicolumn{1}{|c|}{} & \multicolumn{1}{c|}{$\tilde\gamma_2+\tilde\gamma_3<0$} & \multicolumn{1}{c|}{$[0 0 0]\rightarrow\{[0 0 1]\}_2$} & \multicolumn{1}{c|}{} & \multicolumn{1}{c|}{} & \multicolumn{1}{c|}{} \\ \hline

\multicolumn{6}{l}{\footnotesize{$^{a}$If $\frac{2u_p}{2u_p+v_p}\in \left(\frac{2\tilde\gamma_1}{2\tilde\gamma_1+\tilde\gamma_2+\tilde\gamma_3},+\infty \right)$.} \footnotesize{$^{b}$If $\frac{2u_p}{2u_p+v_p}\in \left(-\infty,\frac{2\tilde\gamma_1}{2\tilde\gamma_1+\tilde\gamma_2+\tilde\gamma_3} \right)$.} \footnotesize{$^{c}$If $\frac{u_p+v_p}{u_p}\notin(\frac{2\tilde\gamma_1}{2\tilde\gamma_1+\tilde\gamma_2+\tilde\gamma_3},0)$.} \footnotesize{$^{d}$If $\frac{u_p+v_p}{u_p}\in(\frac{2\tilde\gamma_1}{2\tilde\gamma_1+\tilde\gamma_2+\tilde\gamma_3},0)$.}}\\

\end{tabular}
\caption{Non-equilibrium ferroelectric phase evolution under different conditions for light  circularly polarized in the $x-y$ plane. The symbol $\{[u v w]\}_n$ denotes the $Z_n$ symmetry breaking phase with polarization along $u\hat{x}+v\hat{y}+w\hat{z}$ or the other $n-1$ equivalent directions related by $C_4$ rotations about the $z$ axis or reflection in the  $xy$ plane ($z\to-z$).
}
\label{CircP}

\begin{tabular}{cccccc}
\hline
\multicolumn{2}{|c|}{\multirow{2}{*}{}} & \multicolumn{2}{c|}{$\tilde\gamma_1>0$} & \multicolumn{2}{c|}{$\tilde\gamma_1<0$} \\ \cline{3-6} 
\multicolumn{2}{|c|}{} & \multicolumn{1}{c|}{$\tilde\gamma_1+\tilde\gamma_2+\tilde\gamma_3>0$} & \multicolumn{1}{c|}{$\tilde\gamma_1+\tilde\gamma_2+\tilde\gamma_3<0$} & \multicolumn{1}{c|}{$\tilde\gamma_1+\tilde\gamma_2+\tilde\gamma_3>0$} & \multicolumn{1}{c|}{$\tilde\gamma_1+\tilde\gamma_2+\tilde\gamma_3<0$} \\ \hline
\multicolumn{1}{|c|}{\multirow{2}{*}{$v_p>0$}} & \multicolumn{1}{c|}{$\tilde\gamma_2+\tilde\gamma_3<0$} & \multicolumn{1}{c|}{\begin{tabular}[c]{@{}c@{}}$[0 0 0]\rightarrow \{[0 1 1]\}_4$$^{a}$\\ $[0 0 0]\rightarrow\{[0 1 1]\}_4\rightarrow \{[a b b]\}_8$$^{b}$\end{tabular}} & \multicolumn{1}{c|}{\multirow{2}{*}{\begin{tabular}[c]{@{}c@{}}$[0 0 0]\rightarrow\{[0 1 1]\}_4$$^{a}$\\ $[0 0 0]\rightarrow\{[0 1 1]\}_4\rightarrow\{[a b b]\}_8$$^{b}$\end{tabular}}} & \multicolumn{1}{c|}{\multirow{4}{*}{\begin{tabular}[c]{@{}c@{}}$[0 0 0]\rightarrow\{[1 0 0]\}_2$$^{c}$\\ $[0 0 0]\rightarrow\{[1 0 0]\}_2\rightarrow\{[a b b]\}_8$$^{d}$\end{tabular}}} & \multicolumn{1}{c|}{\multirow{4}{*}{$[0 0 0]$}} \\ \cline{2-3}
\multicolumn{1}{|c|}{} & \multicolumn{1}{c|}{$\tilde\gamma_2+\tilde\gamma_3>0$} & \multicolumn{1}{c|}{$[0 0 0]\rightarrow\{[1 0 0]\}_2\rightarrow\{[a b b]\}_8$} & \multicolumn{1}{c|}{} & \multicolumn{1}{c|}{} & \multicolumn{1}{c|}{} \\ \cline{1-4}
\multicolumn{1}{|c|}{\multirow{2}{*}{$v_p<0$}} & \multicolumn{1}{c|}{$\tilde\gamma_2+\tilde\gamma_3<0$} & \multicolumn{1}{c|}{$[0 0 0]\rightarrow\{[0 1 0]\}_4$} & \multicolumn{1}{c|}{\multirow{2}{*}{$[0 0 0]\rightarrow\{[0 1 0]\}_4$}} & \multicolumn{1}{c|}{} & \multicolumn{1}{c|}{} \\ \cline{2-3}
\multicolumn{1}{|c|}{} & \multicolumn{1}{c|}{$\tilde\gamma_2+\tilde\gamma_3>0$} & \multicolumn{1}{c|}{$[0 0 0]\rightarrow\{[1 0 0]\}_2$} & \multicolumn{1}{c|}{} & \multicolumn{1}{c|}{} & \multicolumn{1}{c|}{} \\ \hline
\multicolumn{6}{l}{\footnotesize{$^{a}$If $\frac{2u_p}{2u_p+v_p}\in \left(\frac{\tilde\gamma_1+\tilde\gamma_2+\tilde\gamma_3}{\tilde\gamma_1},+\infty \right)$.} \footnotesize{$^{b}$If $\frac{2u_p}{2u_p+v_p}\in \left(-\infty,\frac{\tilde\gamma_1+\tilde\gamma_2+\tilde\gamma_3}{\tilde\gamma_1} \right)$.} \footnotesize{$^{c}$If $\frac{u_p+v_p}{u_p}\notin(\frac{\tilde\gamma_1+\tilde\gamma_2+\tilde\gamma_3}{\tilde\gamma_1},0)$.} \footnotesize{$^{d}$If $\frac{u_p+v_p}{u_p}\in(\frac{\tilde\gamma_1+\tilde\gamma_2+\tilde\gamma_3}{\tilde\gamma_1},0)$.}}\\

\end{tabular}
\caption{Non-equilibrium ferroelectric phase evolution under different conditions when light is linearly polarized in $x$-direction. The notation is same as that in Table \ref{CircP} except that different $Z_n$ symmetry breaking phases are related by $C_4$ rotation around $x$ axis and reflection with respect to $yz$ plane ($x\to-x$).}
\label{LineP}
\end{table*}

Here we assume the system is homogeneous.  Since the typical wavelength of THz/IR light ($10^1-10^2\mu$m) is much larger than the relevant microscopic scales, we restrict our attention to the uniform response of the $P$ and the $Q$ modes.  The equations of motion (EOMs) of the $Q$ modes and $P$ modes are then given by
\begin{multline}
   \Ddot{Q}_i+\omega_Q^2 Q_i+\frac{\partial V_4(Q_i,P_i)}{\partial Q_i}+\frac{\partial V_{PQ}(Q_i,P_i)}{\partial Q_i}\\+\frac{\partial V_C(Q_i,P_i)}{\partial Q_i}-Z_q E_i=0, \label{FullEOMQ}
\end{multline}
\begin{multline}
    \Ddot{P}_i+\omega_P^2 P_i+\frac{\partial V_4(Q_i,P_i)}{\partial P_i}+\frac{\partial V_{PQ}(Q_i,P_i)}{\partial P_i}\\+\frac{\partial V_C(Q_i,P_i)}{\partial P_i}-Z_p E_i=0, \label{FullEOMP}
\end{multline}

\noindent
with potentials  
\begin{equation}
    V_4=u_q(\sum_i Q_i^2)^2+v_q\sum_i Q_i^4+u_p(\sum_i P_i^2)^2+v_p\sum_i P_i^4,
\end{equation}
\begin{equation}
    V_{PQ}=-\frac{\gamma_1}{2}|\Vec{P}|^2|\Vec{Q}|^2{-}\frac{\gamma_2}{2}(\Vec{P}\cdot\Vec{Q})^2{-}\frac{\gamma_3}{2}\sum_i P_i^2Q_i^2,
\end{equation}
\begin{multline}\label{CoulombV}
    V_C=\sum_{i,j}\left[2\pi Z_p^2\frac{k_i k_j}{k^2} P_i P_j+2\pi Z_q^2\frac{k_i k_j}{k^2} Q_i Q_j\right.\\\left.+4\pi Z_pZ_q\frac{k_i k_j}{k^2}Q_i P_j\right],
\end{multline}
where we implicitly take the long-wavelength limit $\vec{k}\rightarrow 0$ and assume a simple harmonic drive $E_i(t)=E_{0,i}\cos{\Omega t}$.  Several approximations are needed to simplify \eqref{FullEOMQ} and \eqref{FullEOMP}. First, we assume that the high frequency $Q$ modes are not influenced by any feedback from the low frequency $P$ modes ($\omega_Q\gg\omega_P$), so that  Eq. (\ref{FullEOMQ}) becomes
\begin{equation}
    \Ddot{Q}_i+\omega_Q^2 Q_i+4\pi Z_q^2\frac{k_i k_j}{k^2}Q_j+4u_q|\vec{Q}|^2Q_i+4v_q Q_i^3=Z_qE_i,
    \label{eq:Qmode}
\end{equation}
where we have used a summation convention over the repeated subscripts $j$.
Assuming the drive to be weak enough to ignore the cubic terms in Eq. \eqref{eq:Qmode}, we find that the high-frequency transverse modes are then directly proportional to the driving field, \begin{equation}\label{Qchi}
Q_{i}(t) = \chi_q E_{0,i} \cos\Omega t,
\end{equation}
where the susceptibility
\begin{equation}\label{Qsusc}
\chi_q=\frac{Z_q}{\omega_Q^2-\Omega^2}.
\end{equation}
 diverges as $\Omega\rightarrow \omega_Q$, reflecting the resonant response of the $Q$ mode to the driving field. 
 Note that the incoming laser beam contains purely transverse fields, so that longitudinal Q modes are not excited (see discussion after  Eq. \eqref{eq:extfield}). This linear approximation reduces Eq. (\ref{FullEOMP}) to a decoupled nonlinear differential equation for $\vec{P}$ only. 

\begin{figure}[b]
     \centering
     \includegraphics[width=0.48\textwidth]{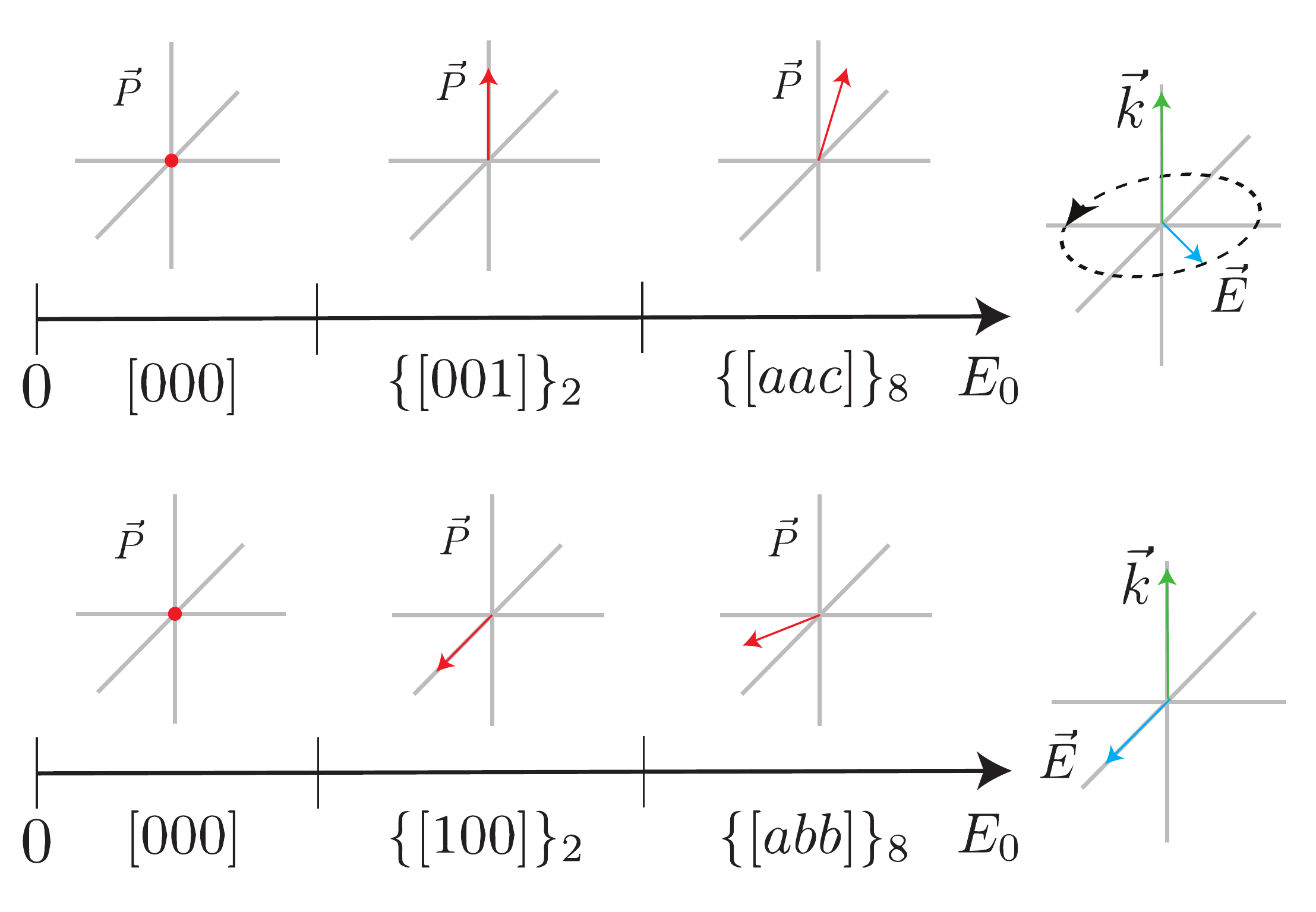}
     \caption{Schematics of two illustrative examples for the nonthermal pathways considered:
      multiple polar phase transitions as a function of electric field amplitude in cubic paraelectric systems. 
      Top: the light is circularly polarized in the $x-y$ plane with $v_p>0$, $\tilde\gamma_1>0$, $\tilde\gamma_2+\tilde\gamma_3<0$ and $2\tilde\gamma_1+\tilde\gamma_2+\tilde\gamma_3>0$; bottom: the light is linearly polarized along the $x$-axis with $v_p>0$, $\tilde\gamma_1>0$, $\tilde\gamma_2+\tilde\gamma_3>0$ and $\tilde\gamma_1+\tilde\gamma_2+\tilde\gamma_3>0$.}
     \label{fig:twoPT}
 \end{figure}

Next, we assume the solution is rapidly oscillating with frequency $\sim\Omega$ around a time-averaged value $\overline{P_{i}}$,
where $\overline{(...)}$ denotes the time-average over a time interval $\tau\gg 1/\Omega$.
The equations for $P_{i}$ are then obtained by time-averaging Eq. (\ref{FullEOMP}) with respect to $Q$s.
The resulting equations are identical to a particle moving in the effective potential:
\begin{align}\label{Veff}
    V_{\text{eff}}&=\sum_{i,j}\left(\frac{\omega_P^2}{2}\delta_{i j}+2\pi Z_p^2\frac{k_i k_j}{k^2}\right)P_i P_j\\\nonumber
    &+u_p(\sum_i P_i^2)^2+v_p\sum_i P_i^4\\\nonumber
    &-\sum_{i,j}\left[\frac{\gamma_1}{2}\overline{Q_j^2}P_i^2+\frac{\gamma_2}{2} \overline{Q_i Q_j}P_i P_j+\frac{\gamma_3}{2} \overline{Q_i^2}P_i^2\right]\\\nonumber
    &\equiv\frac{1}{2} (\tilde \omega_P^2)_{ij}P_i P_j + u_P( \sum_i P_i^2)^2 + v_P \sum_{i}P_{i}^4,
\end{align}
where we have employed a summation convention in the final expression. 
Effective potentials of this sort lie at the heart of light-matter manipulations, and $V_{\rm eff}$ is in essence, a simple extrapolation of the laser-tweezer concept to a many-body phonon potential. We note that using \eqref{Qchi} we can rewrite the resonant response of $Q_\alpha = \chi_q E_\alpha$ allowing us to rewrite the effective potential in terms of the driving field
\begin{eqnarray}
\label{VeffQ}
V_{\text{eff}}&=&\sum_{i,j}\left(\frac{\omega_P^2}{2}\delta_{ij}+2\pi Z_p^2\frac{k_i k_j}{k^2}\right)P_i P_j\cr
    &+&u_p(\sum_i P_i^2)^2+v_p\sum_i P_i^4\cr
    &-&\sum_{i,j}\left[\frac{\tilde{\gamma}_1}{2}\overline{E_j^2}P_i^2+\frac{\tilde{\gamma}_2}{2} \overline{E_i E_j}P_i P_\beta+\frac{\tilde{\gamma}_3}{2} \overline{E_i^2}P_i^2\right]
\end{eqnarray}
where the coefficients $\tilde\gamma_i = (\chi_q)^2\gamma_i$ ($i=1,2,3$) are the resonant response coefficients to the external field.  
We note that in steady state the time-average $\overline{{P}_{i}}$ is expected to lie at the local minimum of $V_{\rm eff}$. 

We now analyze the instabilities resulting from $V_{\rm eff}$  in the presence of a circularly or linearly polarized electromagnetic wave propagating along the $z$ axis, exciting the transverse modes $Q_x$ and $Q_y$. We assume that the LO-TO splitting $\sim Z_p^2$ is large and only consider the long-wavelength soft transverse phonon modes, for instance e.g. $P_{x,y}(0,0,k_z)$ and $P_{z}(k_x,0,0)$. The excitation of $Q$ modes generates an anisotropic shift in the effective frequency of the soft transverse $P$ phonons. They are defined by the eigenvalues of second derivative matrix of (\ref{Veff}), given by
\begin{eqnarray}\label{freqshift}
    &(\tilde{\omega}_P^2)_{xx}=\omega_P^2-\tilde\gamma_1(\overline{E_x^2}+\overline{E_y^2})-(\tilde\gamma_2+\tilde\gamma_3)\overline{E_x^2},\\\nonumber
&(\tilde{\omega}_P^2)_{yy}=\omega_P^2-\tilde\gamma_1(\overline{E_x^2}+\overline{E_y^2})-(\tilde\gamma_2+\tilde\gamma_3)\overline{E_y^2},\\ \nonumber
&(\tilde{\omega}_P^2)_{zz}=\omega_P^2-\tilde\gamma_1(\overline{E_x^2}+\overline{E_y^2}),
\end{eqnarray}
Note that we have used $\overline{E_xE_y}=0$. 
For circularly polarized light running along the $z$ axis $\overline{E_x^2}=\overline{E_y^2}=\frac{1}{2}E_{0,x}^2=\frac{1}{2}E_{0,y}^2$, and then Eq. \eqref{freqshift} becomes
 \begin{eqnarray}\label{example}
 (\tilde{\omega}_P^2)_{xx}= (\tilde{\omega}_P^2)_{yy}& = & \omega_P^2 - (2\tilde\gamma_1+\tilde\gamma_2+\tilde\gamma_3)\overline{E_x^2}, \cr
     (\tilde{\omega}_P^2)_{zz}&=&\omega_P^2-2\tilde\gamma_1 \overline{E_x^2}.
     \label{eq:hessian_circ_1}
 \end{eqnarray}
 Suppose that both  $\tilde\gamma_1>0$ and $2\tilde\gamma_1+\tilde\gamma_2+\tilde\gamma_3>0$  are positive.  If $\tilde\gamma_2+\tilde\gamma_3>0$, then as the magnitude of drive increases, the transverse mode frequencies $(\tilde{\omega}_P^2)_{xx}=(\tilde{\omega}_P^2)_{yy}$ vanish first, giving rise to a spontaneous polarization in the $x-y$ plane in the steady state once $\overline{E_{x(y)}^2}>\omega_p^2/(2\tilde\gamma_1+\tilde\gamma_2+\tilde\gamma_3)$. The direction of the polarization that develops is determined by the anisotropy constant $v_p$. From Eq. (\ref{Veff}), one finds that the effective potential $V_{\text{eff}}$ is minimized by $\vec{P}$ along $[\pm 1 1 0]$ if $v_p>0$ and by $\vec{P}$ along $[1 0 0]$ or $[0 1 0]$ if $v_p<0$.

\begin{figure*}[t]
 \includegraphics[width=0.95 \textwidth]{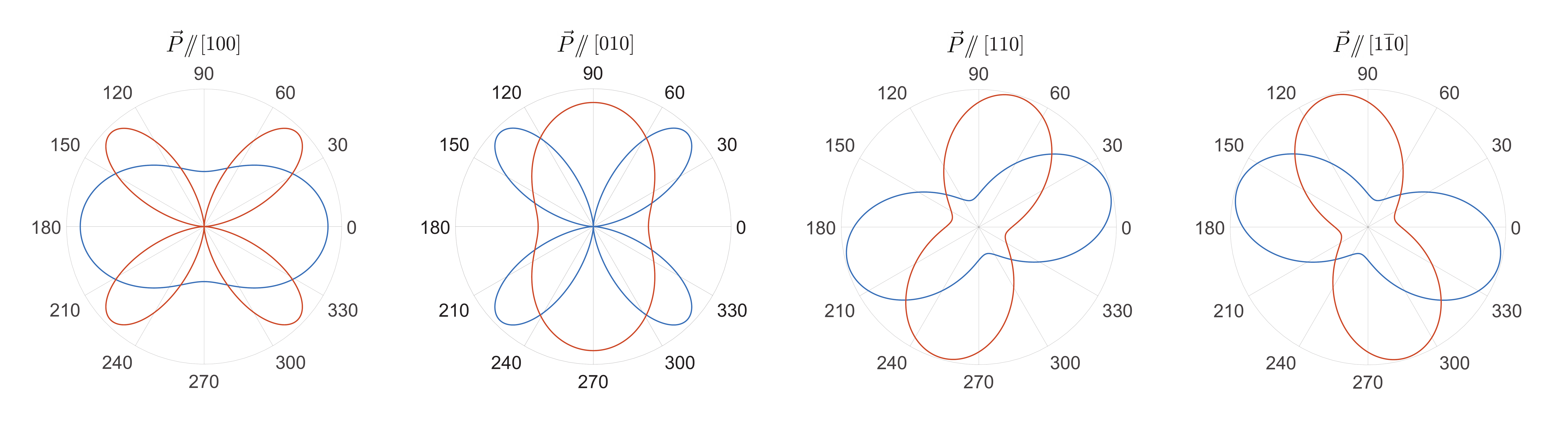}
    \caption{Typical angular dependence of intensity $I_x(\theta)$ (blue) and $I_y(\theta)$ (red) of second harmonic light in laboratory frame for some of the non-equilibrium ferroelectric phases with high symmetry. The linearly-polarized probe pulse propagates along $z$ direction and the angle between light polarization and $x$ axis is $\theta$. For phases with $C_{2v}$ symmetry, we choose the nonlinear susceptibility in the crystal frame to be $\chi^\prime_{zxx}=1$, $\chi^\prime_{zzz}=1.5$ and $\chi^\prime_{xzx}=0.4$. Note that we have rescaled some of the curves to make their magnitudes comparable. }
\label{Ivstheta}
\end{figure*}
 
Let us now consider enhancing the drive fluence beyond the critical one. Note that equation \eqref{eq:hessian_circ_1} no longer determines the phonon frequencies, and the stability of the system is determined by the Hessian matrix at the new energy minimum with nonzero $\vec{P}$. 
Let us focus on the case $v_p>0$. If the drive $\overline{E_{x(y)}^2}$ is increased beyond the first instability threshold, the frequency for the transverse $P_z$ mode around the new minimum will soften at a second critical fluence, if the parameters obey certain constraints (see Table \ref{CircP}). This gives rise to a second phase transition. For light with linear polarization along the $x$-axis, the effective frequency Eq. (\ref{freqshift}) can be similarly determined by setting $\overline{E_y^2}=0$.

 Table \ref{CircP} and Table \ref{LineP} summarize various possible ferroelectric orderings that are possible in the effective potential approximation with circular (Table \ref{CircP} ) and linear (Table \ref{LineP}) light polarizations. There are multiple continuous phase transitions when $\overline{E^2}$ (proportional to the intensity of light), varies. Importantly, in equilibrium only $[111]_8$ (for $v_P>0$) or $[100]_6$ (for $v_P<0$, sixfold degenerate due to cubic symmetry) phases can be realized by tuning $\omega_P^2$. Therefore, our analysis shows that an external drive can induce ferroelectric phases that are inaccessible in equilibrium. 
  In Figure \ref{fig:twoPT} we show illustrative examples of two-stage symmetry-breaking driven by circularly and linearly polarized light that do not occur in thermal polar pathways.
 
\color{black}

\subsection{Second Harmonic Generation Signatures}
\label{SHGsigs}

Experimentally, non-equilibrium ferroelectricity is detected via second harmonic generation (SHG) \cite{nelson2019,cavalleri2019}. Due to the nonlinearity, a monochromatic electric field with frequency $\omega$ induces dipole moments oscillating at a doubled frequency $2\omega$, described by the second-order nonlinear optical susceptibility tensor $\chi$ \cite{boyd2020}
\begin{equation}
    P_i(2\omega)=\chi_{ijk}E_j(\omega)E_k(\omega). \label{SHG}
\end{equation}
The dipole moments then act as a source and generate a second harmonic of frequency $2\omega$ and intensity $I_i(2\omega)\propto |P_i(2\omega)|^2$.

\begin{figure*}[t]
  \subfigure[]{
    \label{fig:ClPvsE} 
    \includegraphics[width=0.32 \textwidth]{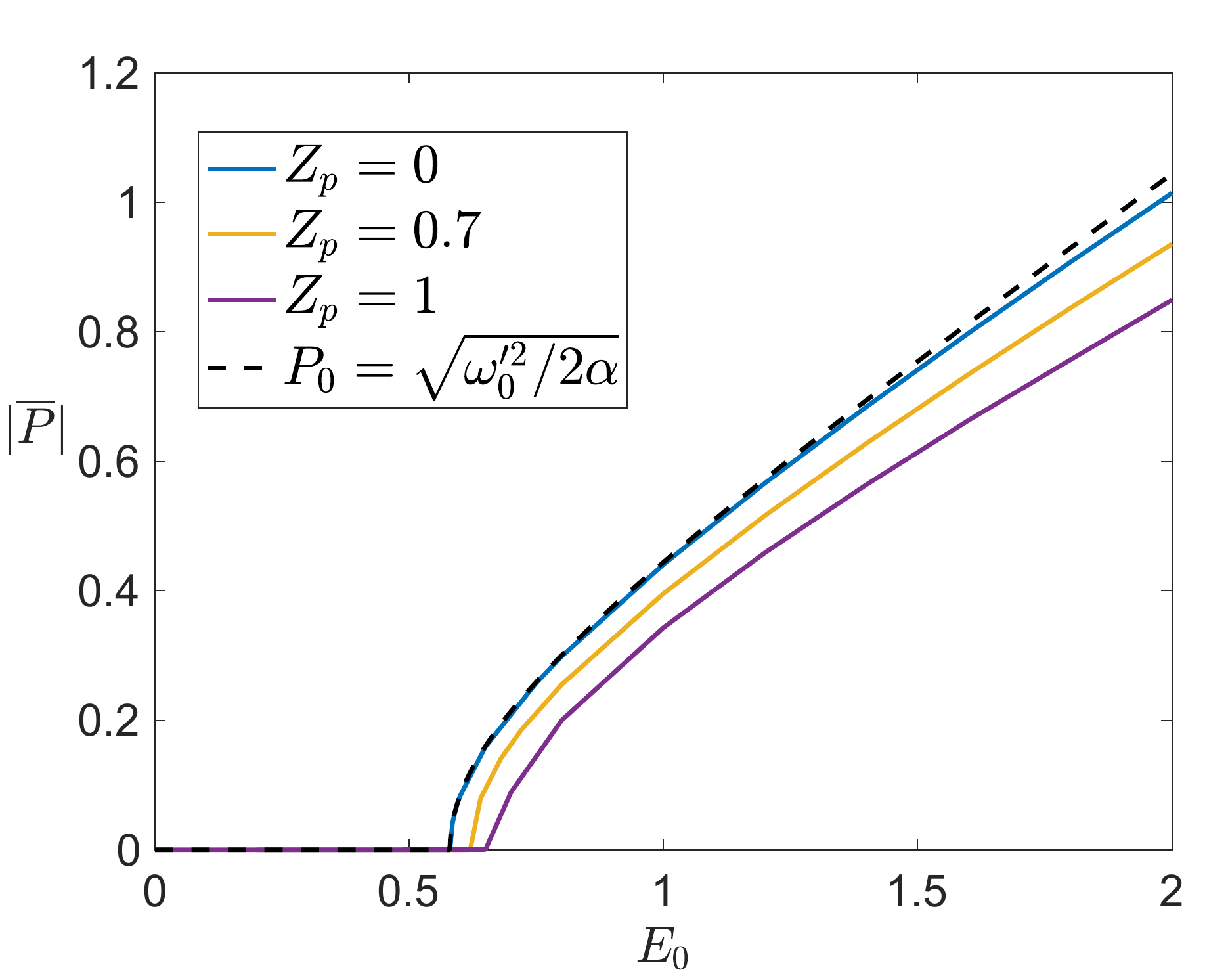}}
    \subfigure[]{
    \label{fig:CLPvst} 
    \includegraphics[width=0.32 \textwidth]{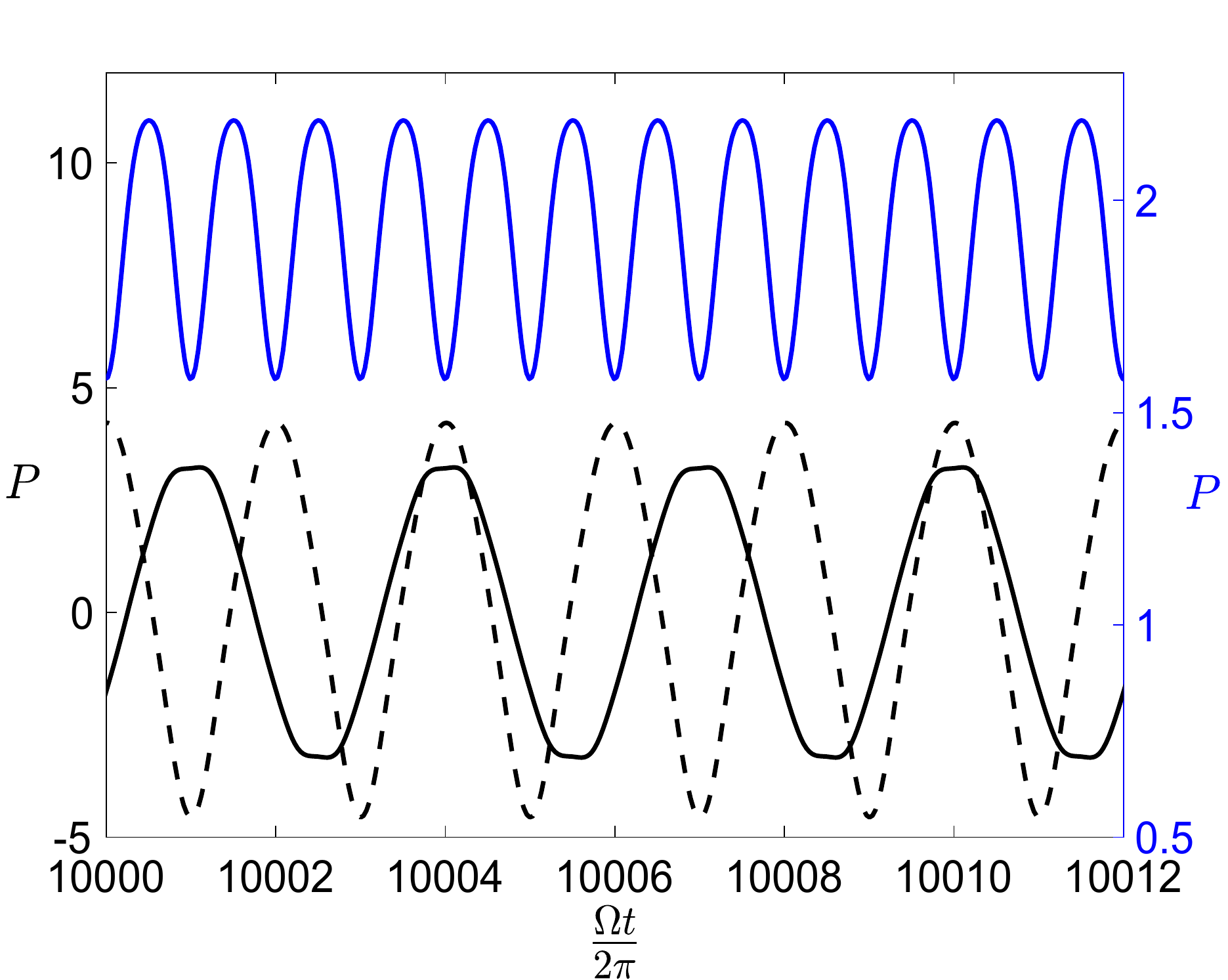}}
          \subfigure[]{
    \label{fig:CLPvsEMesh} 
    \includegraphics[width=0.32\textwidth]{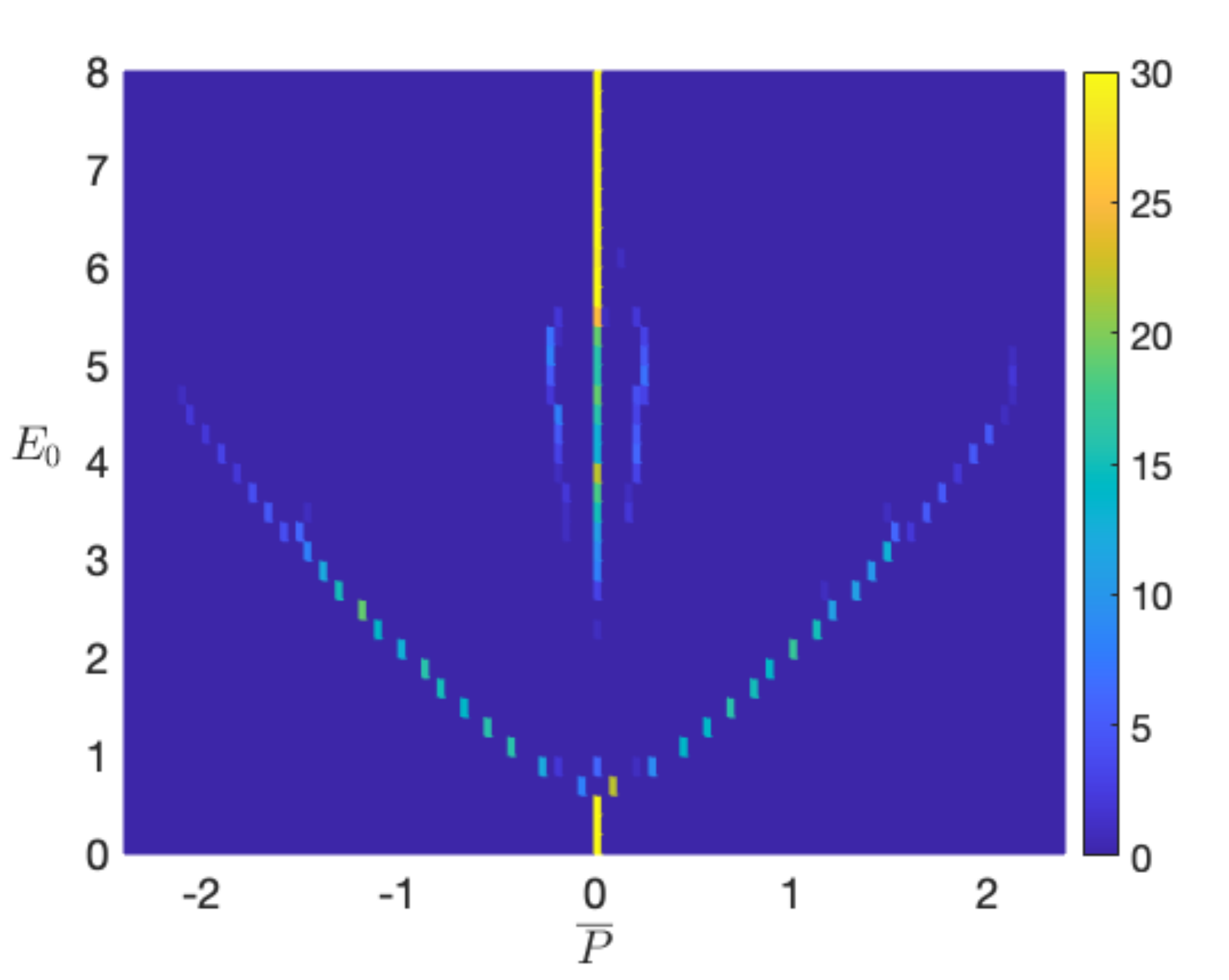}}
    \subfigure[]{
    \label{fig:chaos} 
    \includegraphics[width=0.32\textwidth]{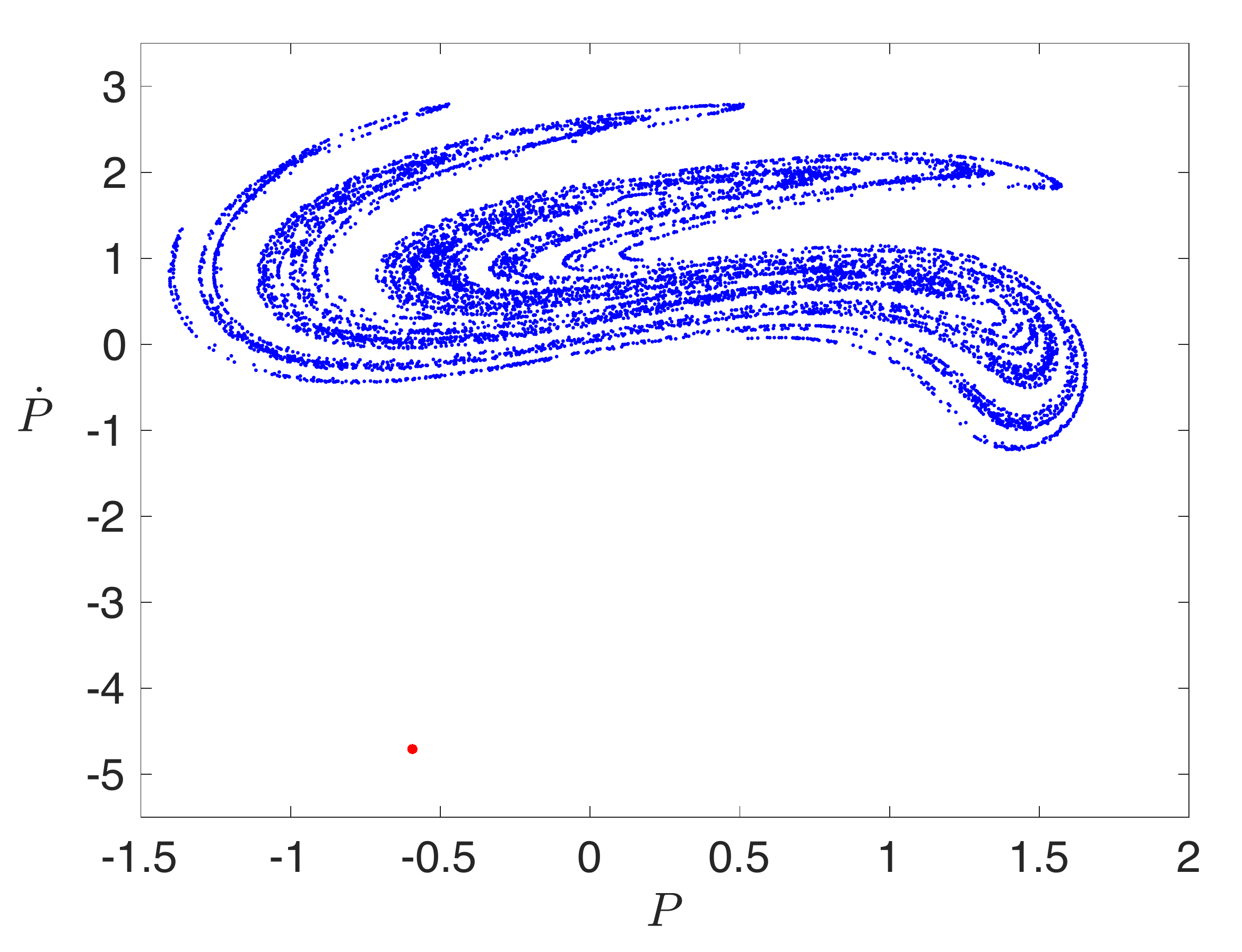}}
    \subfigure[]{
    \label{fig:CLPvst2} 
    \includegraphics[width=0.32\textwidth]{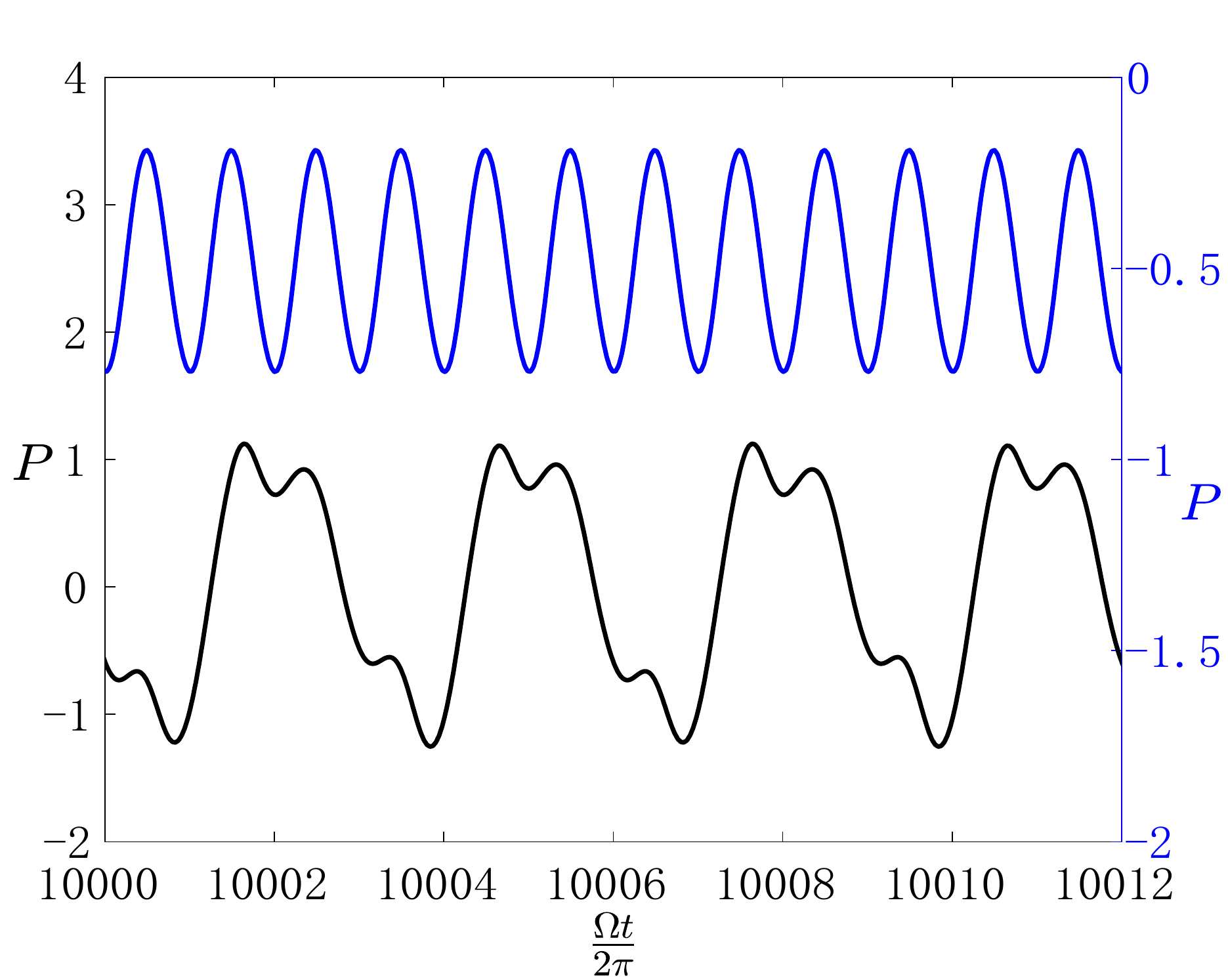}}
    \subfigure[]{
    \label{fig:CLPvsEMesh2} 
    \includegraphics[width=0.32\textwidth]{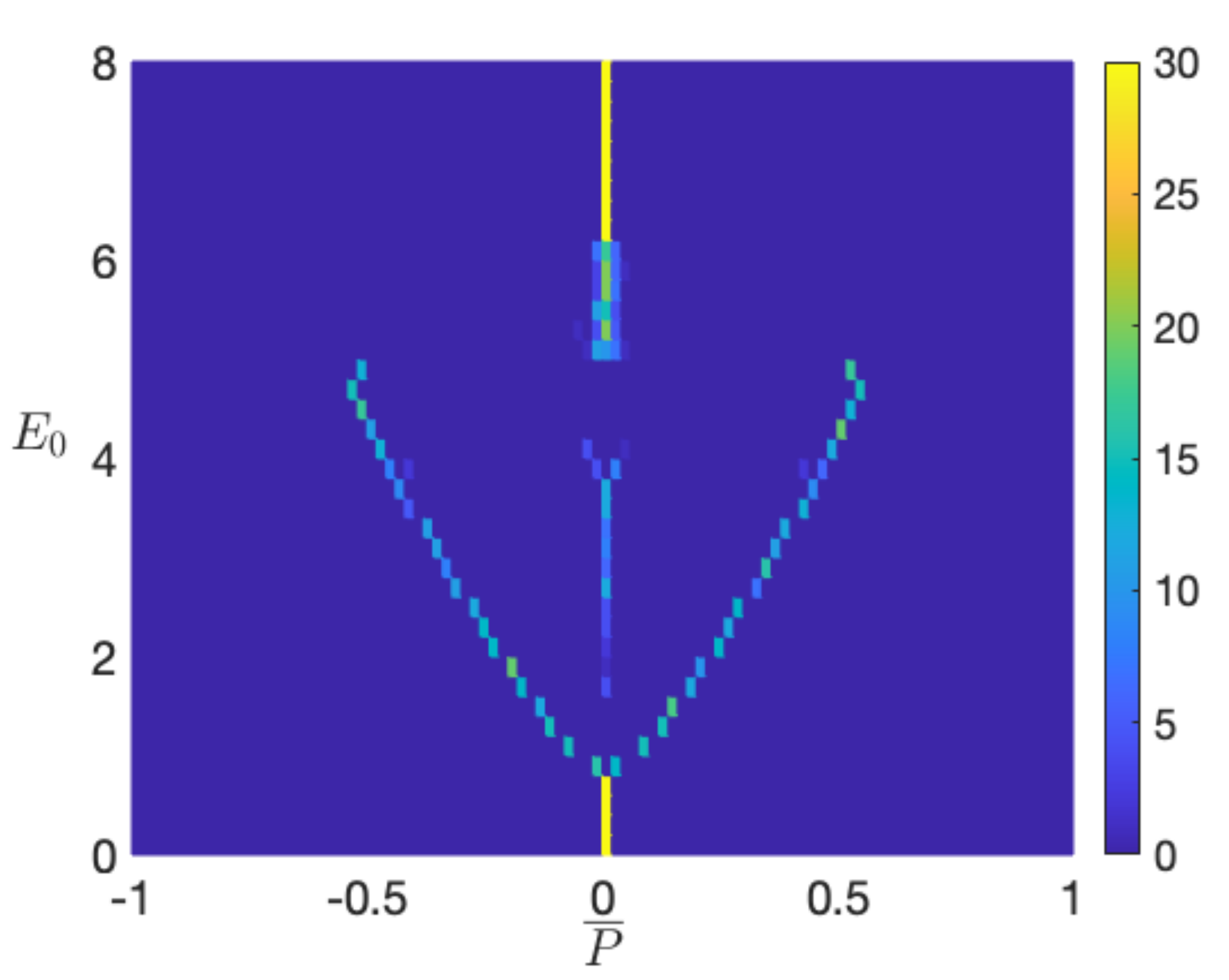}}

    \caption{
    (a): Time-averaged polarization $\overline{P}$ versus electric field strength $E_0$. The solid lines are the numerical solutions of Eq. \eqref{TwoHOmodel} for different $Z_p$, while the dashed line is the analytic solution $P_0=\pm \sqrt{\omega_P^{\prime 2}/2\alpha}$ when time-dependent terms is absent. (b) and (e): Polarization versus time in the steady state when $E_0=4, Z_p=0.3$, obtained by numerical method; the blue line is the regular solution approximately captured by Eq. (\ref{ClSolution}) while the black lines correspond to the solutions with frequency fractional of $\Omega$. (c) and (f): Possible values of $\overline{P}$ versus $E_0$ when $Z_p=0.3$; the result is obtained with $30$ random initial conditions for each fixed $E_0$, and the color of grids indicates the number of times the system reaches $\overline{P}$ in its steady state. (d) The Poincar\'e section when $E_0=5.1, Z_p=0.3$, suggesting the coexistence of chaotic behavior (blue) and periodic solution (red), i.e. KAM structure \cite{lichtenberg2013regular}. All the steady state solution is obtained by numerically solving Eqs. (\ref{EOM:Q}) and (\ref{EOM:P}). Common parameter values for all plots: $\omega_P=0.1, \omega_Q=2, \Omega=2.1, \gamma=0.01, Z_q=1$, For (a)-(c), we choose $\alpha=0.1$, $\beta_1=\beta_0=0.002$; for (d)-(e)  we choose $\alpha=1$,  $\beta_1=\beta_0=0.1$. }
\label{fig:Classical}
\end{figure*}

For centrosymmetric systems, the absence of inversion symmetry breaking causes all elements of $\chi$ vanish, so there is no SHG. For noncentrosymmetric systems, the residual symmetry typically reduces the eighteen independent tensor elements to only a few,  constraining the relation of intensities along different directions. For example, consider the $\{[110]\}_4$ phase listed in Table \ref{CircP},  described by the  effective potential Eq. (\ref{Veff}) and Eq. (\ref{example}).
The polarized incident light breaks the symmetry between $x,y$ and $z$ directions, so   the ferroelectric phase with $\vec{P}=(P_0,\pm P_0,0)$ only has $C_{2v}$ symmetry along the $\hat{n}=[1,\pm1,0]$ axis. For convenience, we denote the symmetry axis $\hat{n}$ as $z^\prime$ and the other two perpendicular directions as $x^\prime$ and $y^\prime$, referred as the crystal frame. For $C_{2v}$ symmetry, there are only five non-zero independent tensor elements, that is $\chi^\prime_{xzx}=\chi^\prime_{xxz}, \chi^\prime_{yyz}=\chi^\prime_{yzy}, \chi^\prime_{zxx}, \chi^\prime_{zyy}, \chi^\prime_{zzz}$ in the crystal frame. Suppose that the material interacts with a probe pulse with linear polarization along $\hat{\theta}^\prime=\cos \theta \hat{z}^\prime+\sin \theta \hat{x}^\prime$ direction, using Eq. (\ref{SHG}) one obtains 
\begin{align}
    P_z^\prime&=E^2(\chi^\prime_{zxx}\sin^2 \theta+\chi^\prime_{zzz}\cos^2 \theta),\\\nonumber
P_x^\prime&=E^2\chi^\prime_{xzx}\sin 2\theta,\\ \nonumber
P_y^\prime&=0
\end{align}
in the crystal frame. Consider $\{[aac]\}_8$ in Table \ref{CircP} as another example. It has $C_{1h}$ symmetry and the mirror plane is perpendicular to $\hat{y}^\prime$. Besides the five nonzero elements in $C_{2v}$, there are other five non-vanishing elements: $\chi^\prime_{xxx},\chi^\prime_{xyy},\chi^\prime_{xzz},\chi^\prime_{yxy}=\chi^\prime_{yyx},\chi^\prime_{zzx}=\chi^\prime_{zxz}$. Thus for probe pulse with polarization along $\hat{\theta}^\prime$, one has
\begin{align}
    P_z^\prime&=E^2(\chi^\prime_{zxx}\sin^2 \theta+\chi^\prime_{zzz}\cos^2 \theta+\chi^\prime_{xzx}\sin 2\theta),\\\nonumber
P_x^\prime&=E^2(\chi^\prime_{xxx}\sin^2\theta+\chi^\prime_{xzz}\cos^2\theta+\chi^\prime_{xzx}\sin 2\theta),\\ \nonumber
P_y^\prime&=0.
\end{align}
Thus one can observe a change of profile $P_i(\theta)$ when there is non-equilibrium phase transition. In Fig. \ref{Ivstheta}, we show the typical profile of $P_i(\theta)$ for some of the high-symmetry phases, which is experimentally measurable \cite{cavalleri2019}.

\subsection{A Minimalist Coupled Oscillator Model} 
\label{TCOM}

We now go beyond the effective potential approximation to study the dynamics of anharmonically coupled oscillators. For simplicity, we consider a minimalist model with  two scalar harmonic oscillators $P$ and $Q$. The real-time action is given by
\begin{align}
    S=\int dt &\left[\frac{1}{2}\dot{Q}^2-\frac{1}{2}\omega_Q^2 Q^2+\frac{1}{2}\dot{P}^2-\frac{1}{2}\omega_P^2 P^2\right.\label{TwoHOmodel}\\\nonumber
    &\left.-\frac{1}{4}\alpha P^4+\frac{\gamma}{2}P^2Q^2+Z_qQE+Z_pPE\right], \\\nonumber
\end{align}
 where $\alpha>0$ and we assume $\omega_Q\ll\omega_P\approx\Omega$. To obtain this model, we have restricted ourselves to the uniform states and neglected insignificant terms to our interests, such as the anharmonic interactions of $Q$ modes. We also neglected linear-linear and linear-cubic couplings between $P$ and $Q$ modes. These couplings lead to terms in the equation of motion oscillating with frequency $\Omega$ and its multiples with zero average, therefore we suggest that their effects should be qualitatively similar to the effect of direct coupling $Z_p$ to the oscillating electric field (see Appendix \ref{app:PQeffects} for more discussions and numerical justification). In particular, the effect of the linear $P-Q$ coupling can be absorbed into the renormalization of the coupling of $P$ mode to light $Z_p$. We leave the detailed study of possible additional effects of cubic-linear couplings (such as higher harmonic driving effects) to future work.
 Note that we are working in the regime where the $Q$ mode is only driven quasi-resonantly where the amplitude of $Q$ mode is not too large and the nonlinearity of $Q$ mode, whose effect has been studied in Ref. \cite{itin2018}, does not play an important role. This simplified model could describe the non-equilibrium PE-FE transition driven by the external electric field, given that the unstable soft phonon mode is non-degenerate near the phase transition, which is true for $Z_2$ symmetry breaking transitions listed in Table \ref{CircP} and Table \ref{LineP}.

\begin{figure}[tb]
    \centering
    \includegraphics[width=0.48\textwidth]{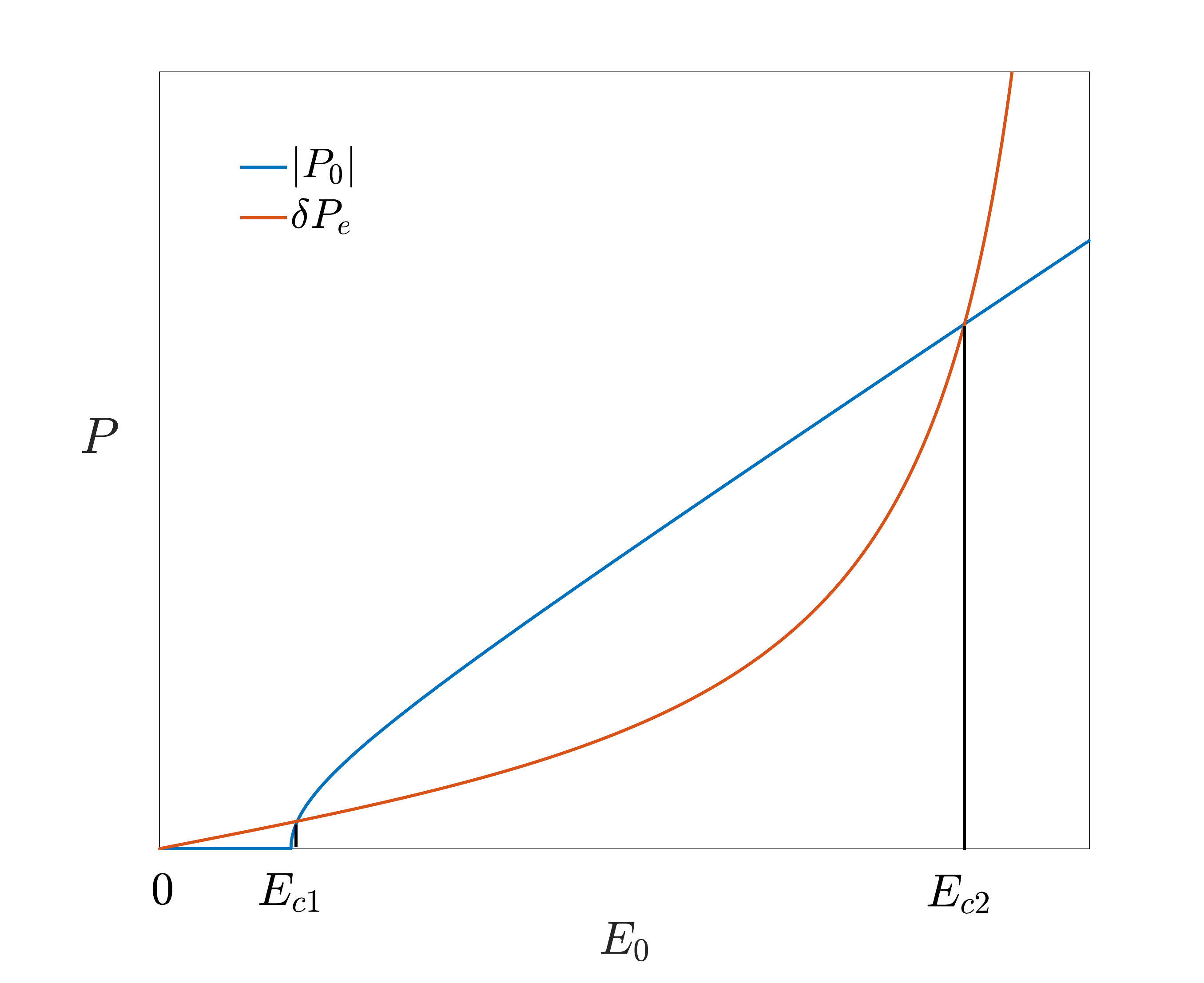}
    \caption{Field dependence of the steady-state polarization $|P_0|$ and the amplitude of rapid oscillations $\delta P_e$ (see \eqref{dpvsp}). For $\delta P_e>|P_0|$, the rapid oscillations can no longer be neglected, having the effect of  reducing the observable steady-state polarization to zero.}
    \label{fig:dpvsp}
\end{figure}


The classical equations of motion (EOMs) resulting from  Eq. (\ref{TwoHOmodel}) are
\begin{align}
    \Ddot{Q}+\omega_Q^2 Q+\beta_1 \dot{Q}-\gamma P^2Q&=Z_q E(t),\label{EOM:Q}\\
    \Ddot{P}+\omega_P^2 P+\beta_0\dot{P}-\gamma Q^2P+\alpha P^3&=Z_p E(t) \label{EOM:P},
\end{align}
where we have added extra phenomenological damping terms $\beta_1\dot{Q}$ and $\beta_0\dot{P}$. Since the $P$ mode is driven off-resonantly by the $E$ field, we expect the magnitude of $P$ to be small so that $\gamma \overline{P^2}\ll \omega_Q^2$, and we may ignore the nonlinear term in Eq. (\ref{EOM:Q}).
Neglecting damping, we obtain the steady state solution 
\begin{equation}
Q(t)=\chi_q E_0 \cos (\Omega t),
\label{Qsteady}
\end{equation}
where
\begin{equation}
\chi_q= Z_q/(\omega_Q^2-\Omega^2)
\end{equation}
is the resonant Q-susceptibility introduced in \eqref{Qsusc}
and Eq. (\ref{EOM:P}) becomes
\begin{equation}
    \Ddot{P}+\left(\omega_P^2+m(t)\right)P+\alpha P^3=Z_p E(t), \label{EOMFull2}
\end{equation}
where $m(t)=-\gamma Q^2(t)$ is the time-dependent mass arising from the $P^2Q^2$ interaction. For convenience, we write the above equation as
\begin{equation}
    \Ddot{P}+\left(\omega_P^2-\gamma \overline{Q^2}\right)P+\alpha P^3=Z_p E(t)+f(t), \label{EOMFull}
\end{equation}
where $\overline{Q^2}=\frac{1}{2}(\chi_qE_0)^2$ is the time-average of the rapidly oscillating $Q$ mode and $f(t)=\frac{1}{2}\gamma \chi_q^2 (E_0)^2 \cos(2\Omega t) P$. When $\omega_P^2-\gamma \overline{Q^2}>0$, the eigenfrequency of the $P$ oscillator remains positive so one expects that $P$ mode oscillates around the global minimum $P=0$. However when $\omega_P^2-\gamma \overline{Q^2}<0$, the mass of the $P$ mode becomes negative and the system becomes ferroelectric in the absence of time-dependent terms, with finite polarization $P_0=\pm \sqrt{\omega_P^{\prime 2}/2\alpha}$ as shown in Fig. \ref{fig:ClPvsE}. Here 
\begin{equation}\label{omp}
(\omega_P^\prime)^2=2(\gamma \overline{Q^2}-\omega_P^2)= 2(\tilde\gamma \overline{E^2}-\omega_P^2),
\end{equation} 
is the eigenfrequency for oscillations in the ferro-electric state and $\tilde\gamma = \gamma \chi_q^2$ is the isotropic version of the resonant response coefficients introduced in \eqref{VeffQ}. In the presence of $E(t)$ and $f(t)$, $P$ periodically oscillates around $P_0$ in the steady state, as seen in Figs. \ref{fig:CLPvst} and \ref{fig:CLPvst2}. We may expand the potential around $P_0$, approximate $P$ in $f(t)$ by $P_0$ in Eq. (\ref{EOMFull}) to obtain the steady state solution
\begin{equation}
    \delta P(t)\approx\frac{Z_pE_0}{\omega_P^{\prime 2}-\Omega^2}\cos(\Omega t)+\frac{\omega_P^{\prime 2}+2\omega_P^{2}}{2(\omega_P^{\prime 2}-4\Omega^2)}P_0\cos(2\Omega t), \label{ClSolution}
\end{equation}
where $\delta P(t)=P-P_0$.  This approximation works well when the maximum oscillation amplitude $\delta P_{e}$ is much smaller than the time-averaged polarization $\overline{P}\approx P_0$, $\delta P_{e}/|P_0|\ll 1$. The term $\delta f(t) \propto \cos(2\Omega t) \delta P$ that we have neglected reduces $|\overline{P}|$, which can be seen by substituting Eq. (\ref{ClSolution}) into $\delta f(t)$ and averaging over time. The  discrepancies between the exact numerical solutions and the approximate solution Eq. (\ref{ClSolution}) displayed in Fig. \ref{fig:ClPvsE} become substantial once $\delta P/|P_0|\sim O(1)$.

In order for the system to exhibit a  macroscopic polarization, the oscillations in the polarization must not exceed the width of the potential well, i.e  the magnitude of oscillation $\delta P_{e}$ \eqref{ClSolution} must be smaller than $|P_0|$, or 
\begin{equation}
    \frac{Z_p E_0}{\Omega^2-\omega_P^{\prime 2}}+\frac{\omega_P^{\prime 2}+2\omega_P^{2}}{2(4\Omega^2-\omega_P^{\prime 2})}\sqrt{\frac{\omega_P^{\prime 2}}{2\alpha}}\lesssim\sqrt{\frac{\omega_P^{\prime 2}}{2\alpha}}.
    \label{eq:20}
\end{equation}
Note that as the two contributions in Eq. (\ref{ClSolution}) are phase coherent and in phase, the total magnitude is simply the sum of its separate parts. For $Z_p=0$ and $\omega_P,\omega_P'\ll \Omega$, we observe that the inequality is always fulfilled, i.e. the system becomes ferroelectric for infinitesimal $\omega_P^{\prime}$.
If $Z_p\neq 0$, we can neglect the second term in Eq. (\ref{eq:20}) 
for sufficiently low $\omega_P^\prime$. The critical fields $E_c$ can then be estimated by solving 
\begin{equation}
    \delta P_e\equiv \frac{Z_p E_c}{\Omega^2-\omega_P^{\prime 2}}=\sqrt{\frac{\omega_P^{\prime 2}}{2\alpha}}\equiv|P_0|. \label{dpvsp}
\end{equation}

Using \eqref{omp} to express ${\omega}_P^{\prime 2}= 2 \tilde \gamma (E_c^2 -E_{c0}^2)$, where $E_{c0}^2=\omega_P^2/\tilde{\gamma}$, we can cast this equation in the dimensionless form
\begin{equation}\label{econd}
\left(\frac{E_Z^2}{E_{c0}^2}\right) x = (x-1) (x - (r^2+1))^2,
\end{equation}
where $x=(E_{c}/E_{c0})^2$ $(1\leq x\leq r^2+1)$, $E_Z^2 = \frac{Z_p^2\alpha}{2 \tilde \gamma^2 \omega_P^2} $ and $r^2 = \frac{\Omega^2}{2\omega_P^2}$. In the limit of small $Z_p$, this gives the limiting values $x=1$ corresponding to the lower critical field $\lim_{Z_p\rightarrow 0} E_{c1} =E_{c0}$ and $x=1+r^2$, corresponding to the higher critical field
\begin{equation}\label{approx}
    \lim_{Z_p\rightarrow 0} E_{c2}= E_{c0}\sqrt{1+ \frac{1}{2}\left(\frac{\Omega^2}{\omega_P^2}\right)}.
\end{equation}
Figure \ref{fig:dpvsp} shows the dependence of $|P_0|$ and $\delta P_e$ on the electric field strength $E_0$. Note that $\delta P_e$ first grows linearly with $E_0$, and then increases superlinearly as the denominator $\Omega^2-\omega_P^{\prime 2}$ decreases due to the hardening of phonon frequency $\omega_P^\prime$. By contrast, $|P_0|$ increases linearly at large $E_0$. Therefore, for $Z_p \lesssim Z_q\frac{\Omega^2}{|\omega_Q^2-\Omega^2|}\sqrt{\frac{\gamma}{2\alpha}}$, Eq. (\ref{dpvsp}) has two solutions as shown in Fig. \ref{fig:dpvsp}, which correspond to the estimated lower and higher critical fields. The ferroelectric phase with nonzero steady-state polarization therefore exists only between these two field values.

For $\omega_p'\ll\Omega$, the system first becomes ferroelectric when $\omega_P^\prime \approx \sqrt{2\alpha}Z_pE_0/\Omega^2$, resulting   in shifted critical values $E_{c1}$, consistent with the numerical results in Fig. \ref{fig:ClPvsE}. For sufficiently large $E_0\gtrsim E_{c2}$, the $P$ oscillator  hops between the two minima as $\delta P_e \gtrsim |P_0|$, leading to a reentrant paraelectric phase. Re-entrant para-electricity has been observed in previous numerical simulations \cite{subedi2014,Subedi2017}.
Indeed, as seen from Figs. \ref{fig:CLPvsEMesh} and \ref{fig:CLPvsEMesh2}, the regular solutions described by Eq. (\ref{ClSolution}) finally disappear at sufficiently large $E_0$. 

Remarkably, at intermediate $E_0$ we also observe other solutions with negligible $\overline{P}$, coexisting with the regular solutions. From Figs. \ref{fig:CLPvst}. and \ref{fig:CLPvst2}, one can see that these multiple solutions may have frequency fractional of $\Omega$. We also find chaotic behavior in certain parameter range. Particularly, choosing different initial conditions, we find two different orbits near the onset of chaos (see Fig. \ref{fig:chaos}). This suggests the coexistence of chaotic behavior with the periodic solution, known as the Kolmogorov–Arnold–Moser (KAM) structure \cite{lichtenberg2013regular}. Indeed we note that  Eq. (\ref{EOMFull}) is the equation of motion of a Duffing oscillator, generalized due to the additional $f(t)$ term, that is known to exhibit period-doubling bifurcation and chaotic behavior in certain parameter 
regimes \cite{Duffing18,strogatz2018,Chatterjee20}. 

It is useful to have an estimate for the critical electric field strength when finite polarization appears and vanishes. We assume $\omega_Q^2=1136.1 \text{meV$\cdot \AA^{-2} \cdot$amu$^{-1}$}$, $ \omega_P^2=1.39 \text{meV$\cdot \AA^{-2} \cdot$amu$^{-1}$}$, $\alpha=206.9 \text{meV$\cdot \AA^{-4} \cdot$amu$^{-2}$} $, $\gamma=11.6 \text{meV$\cdot \AA^{-4} \cdot$amu$^{-2}$}$ and $Z_p\approx Z_q=1.15 e\cdot \text{amu}^{-1/2}$, which are consistent with the parameters for strained KTaO$_3$ \cite{Subedi2017}.  The double well forms when $|Q_c|\sim 0.346\AA\cdot\text{amu}^{1/2}$, which corresponds to the electric field $E_{c0}\sim 2.7$MV/cm if one neglects dissipation and takes $\Omega^2=1200 \text{meV$\cdot \AA^{-2} \cdot$amu$^{-1}$}$. Enforcing condition Eq. (\ref{eq:20}) one obtains a rough estimate of the lower critical electric field $E_{c1}\approx 2.9$MV/cm and the higher critical electric field $E_{c2}\approx 45.6$MV/cm (which compares with the more approximate estimate \eqref{approx}, which gives $E_{c2}\sim 60 $MV/cm).
Note that $E_{c2}$ corresponds to a large amplitude oscillation $|Q|\sim 5.8\AA\cdot\text{amu}^{1/2} $, suggesting the necessity to include nonlinearity of the $Q$ mode. For example, if quartic terms of $Q$ mode are included, such large oscillation amplitudes are suppressed and it may be possible to observe the vanishing of polarization at a higher critical field $E_{c2}$. We note that our predicted electric field strengths are approachable in experiments where light pulses with large peak fields (estimated around 18 MV/cm \cite{cavalleri2019}) are used.

\begin{figure}
    \centering
    \includegraphics[width=0.48\textwidth]{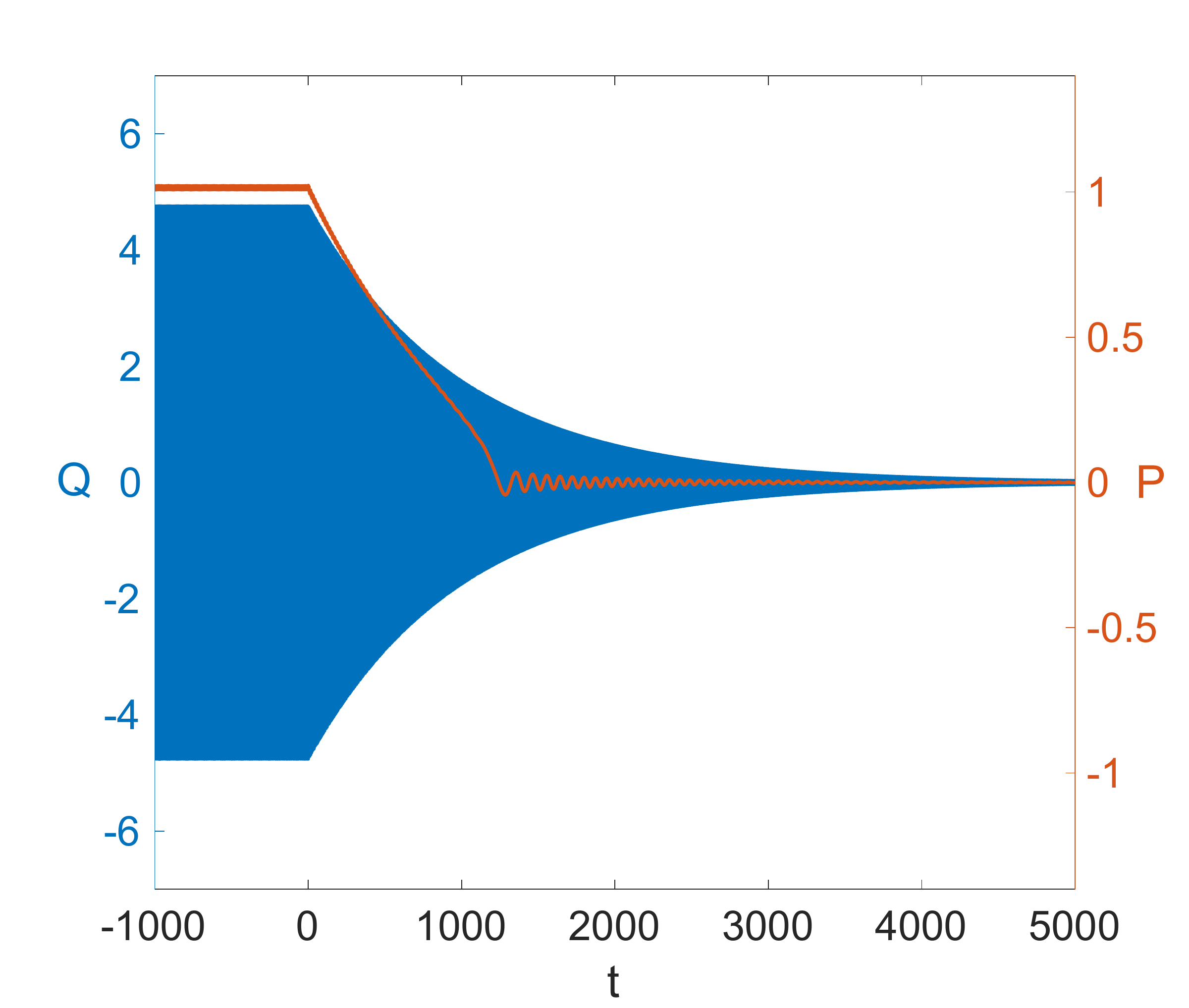}
    \caption{Polarization versus time after the electric field is turned off at $t=0$. We choose $Z_p=0$ and $E_0=2$ when the electric field is on, and set $\beta_1=\beta_0=0.002$ all the time. The rest of parameters are the same as that in Fig. \ref{fig:Classical}.}
    \label{AfterPulse}
\end{figure}

In experiments, the driving is not continuous but is rather performed with finite pulses \cite{nelson2019,cavalleri2019} and the non-equilibrium polarization persists after the pump has been turned off.
In our model, after the external field is turned off, the excited $P$ and $Q$ mode relax due to dissipation resulting
in the decay of polarization. As shown in Fig. \ref{AfterPulse}, after the electric field is off, the amplitude of the
$P$ mode decays but remains finite, until it eventually oscillates around its equilibrium position. This can be understood by noticing that the effective potential felt by $P$ mode gradually relaxes to its equilibrium form due to the damped motion of $Q$. The $P$ mode oscillates around the instantaneous minimum of the effective potential, which becomes zero once $Q^2$ becomes smaller then the critical value for the steady-state driven ferroelectricity. Therefore, the system keeps its `memory' of the pump-induced order for times of order of the $Q$ mode lifetime. This qualitatively describes the observed
persistence of the polarization after the pump is removed \cite{nelson2019,cavalleri2019}, though persistance time-scales
have been reported that are longer than what is accessible in our approach \cite{cavalleri2019}.

\section{Quantum effects in the phase transition}
\label{sec:quant}

We now move to a consideration of quantum effects in light-driven ferroelectricity. 
Previous approaches \cite{gagel2014,Gagel15,mitra2015} focused on
quenches close to a QCP. 

Here we develop a formalism appropriate to our situation where quantum fluctuations coexist with significant classical ones.
\color{black}
In \ref{app:saddle} we demonstrate how the classical equations of motion, Eq. \eqref{EOMFull2}, arise from a quantum Keldysh action for the case of a single nonlinear oscillator and study the quantum corrections in \ref{Q_correction}, where generalizations to include the momentum dispersion of phonons are discussed. 
 The shift of the critical point due to quantum mass corrections is determined as a function of pumping rate, and this should be accessible in experiment.

{ We start with the reduced model
for a soft phonon mode $P(t)$ described in Eq. \ref{EOMFull2},
with effective Lagrangian
\begin{equation} 
{\cal L}[P]= 
\frac{\dot{P}^2}{2}-(\omega_P^2+m(t)) \frac{P^2}{2}
   -\frac{\alpha }{4} P^4+ P E (t).
\end{equation}
The time-dependent $m(t)$ describes the effect of driving on the $P$
modes: we recall that the drive excites the fast $Q$ modes which, within
a classical description, modifies the mass of the slow $P$ mode via
biquadratic interactions. The $P$ mode is also linearly coupled to the
(classical) electric field, $E(t)$; for the purpose of discussion, we have set the
effective charge in \eqref{EOMFull2} to one,
$Z_{p}=1$.

We now quantize this description, describing how we can formulate a
path-integral diagramatic approach. 
The quantum Hamiltonian  is
\begin{equation} 
\hat H = \frac{\hat \pi^{2}}{2}+ 
(\omega_{P}^{2}+m (t))
\frac{\hat P^{2}}{2}
 + \frac{\alpha }{4}\hat P^{4}- \hat P E (t)
\end{equation}
where $\hat \pi $ is the canonical momentum, satisfying
$[\hat P, \hat \pi]= i\hbar  $. 
We now adopt a
Schwinger-Keldysh approach
%
 \begin{figure}[t]
   \centering
  \includegraphics[width=0.48\textwidth]{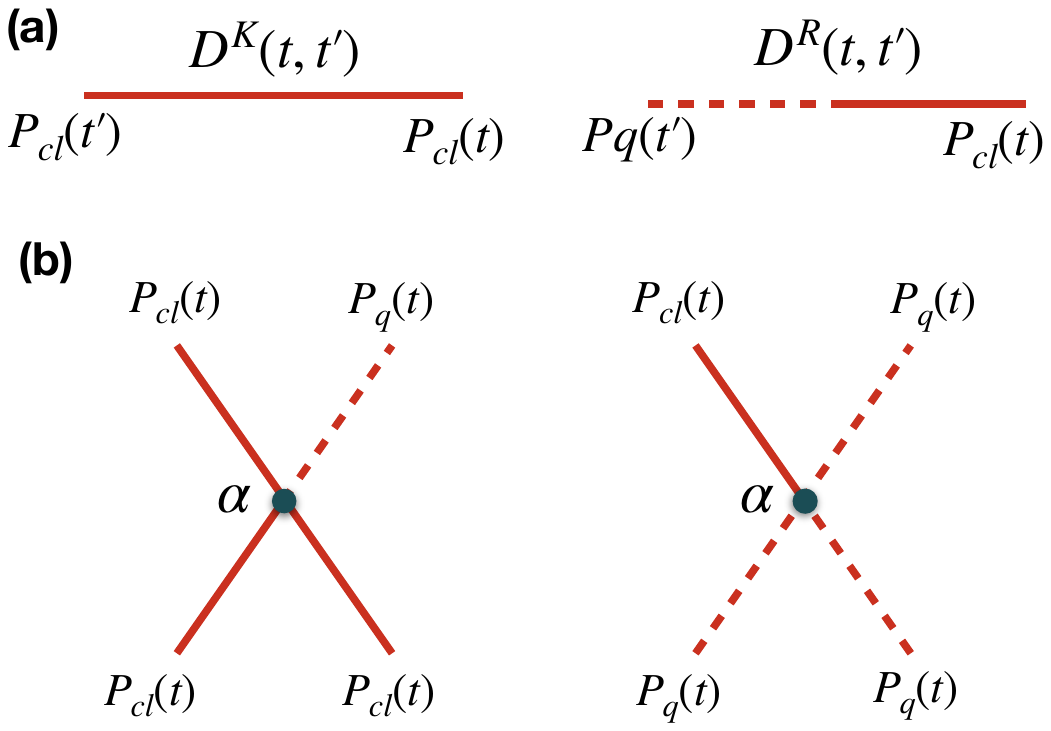}
  \caption{(a) shows graphical representation of the Green's functions in Keldysh field theory consisting of two kinds of fields, namely the classical field $P_{cl}(t)$ (solid line) and the quantum fields $P_q (t)$ (dashed line).  Two independent Green's functions, namely the retarded $D^R(t,t')$ and the Keldysh $D^K(t,t')$ Green's functions, are constructed out of the classical and the quantum fields via Eq.\ref{Green_defn}.  (b) The non-linearity of the $P$ oscillator mode is shown diagrammatically by the quartic interaction vertices with the coupling strength $\alpha$. The interaction vertices are odd in the quantum fields required by the causality structure of the two-contour field theory.} 
   \label{green_vertex}
   \end{figure}
considering a time-evolution from a state of thermal equilibrium in
the distant past.  The generating function $Z_{K}[E]$
is written as 
a time-ordered exponential of the Hamiltonian
over the Keldysh contour in time,  $\{
{\cal C}: t \in  -\infty  \longrightarrow \infty \longrightarrow -\infty \}$
running from the past out to the future and back, 
\begin{equation} 
Z_{K} [E] = {\rm Tr}\left[\hat \rho_{0} {\cal U}_{C}
\right]
\end{equation}
where $\hat \rho_{0}=e^{-\beta  \hat H_0} $ is the initial thermal density matrix {(we will take $T=0$ in our final results)}, while 
\begin{equation} 
U_{C}= U_{-\infty ,\infty }U_{\infty ,-\infty } = T_{C}e^{- \frac{i}{\hbar }\int_{C}dt H (t)},
\end{equation}
where $T_{C}$ denotes path-ordering along the contour ${\cal C}$.
$Z_{K} [E]$ is recast 
as a path integral, 
\begin{eqnarray} 
Z_{K}[E] 
=\int_{\cal C} {D[P]} e^{i {{\cal S}_{K}}/{\hbar }}.
\end{eqnarray}
The action ${\cal S}_{K}$
divides  into 
contributions from the outward and return paths, 
\begin{eqnarray} 
{\cal S}_{K} = \int_{{\cal C}} dt {\cal L}[P]= 
\int_{-\infty }^{\infty }dt \biggl[{\cal L}[P_{+ }]-{\cal L}[P_{- }]\biggr],
\end{eqnarray}
where  
$P_{+} (t)$ and $P_{-} (t)$ 
are the integration variables  on the outward 
and return paths, respectively
\cite{kamenevbook}. 
Under the physical Heisenberg equations of motion, there is strictly
one quantum operator $\hat P (t)$ 
and source term $E(t)$ at each point in time,
but the Keldysh path integral explores the
paths on the upper and lower contours independently, and a complete
generating function must consider independent sources 
$E_{\pm } (t)$ on the
outward and return contours, setting $E_{\pm } (t)=E (t)$ to recover
the physical expectation values. 

Variations of the generating function $Z_{K}(E)$
with respect to the source field $E$ 
generate  correlation functions of the quantum operators $\hat P (t)$, 
path-ordered along the Keldysh contour\cite{Schwinger1951}, such that 
\begin{equation} 
-i\hbar \frac{\delta }{\delta E (t)}\longrightarrow \hat  P (t)
\end{equation}
For instance, 
\begin{eqnarray} 
 \frac{(-i \hbar )^{2}}{Z} \frac{\delta^{2} Z }{\delta E (t_{2})\delta
 E (t_{1})}&\rightarrow & 
\frac{1}{Z}\int {\cal
D}[P]P (t_{1})P (t_{2})e^{i {\cal S}_{K}/\hbar}
\cr
&=& \langle T_{C} \hat P (t_{1})\hat P (t_{2}) \rangle  
\end{eqnarray}
}


{ We adopt a classical-quantum basis
 \begin{eqnarray}
P_{cl}(t)&=&(P_+ + P_-)/2,\cr
P_{q}(t)&=&(P_+ - P_-),
\end{eqnarray}
where the classical and
quantum variables, $P_{cl}$ and $P_{q}$ respectively, are analagous to the 
the center of mass, and relative co-ordinates of two body dynamics. Note that our notation differs from \cite{altlandbook,kamenevbook} by a factor of two in $P_q$, which simplifies some of the intermediate calculations, but without affecting the final results.
} 
The connectivity of the forward and backward paths causes the
joint Green's function $\langle P_q(t)P_q(t')\rangle =0$ to vanish, 
leaving two independent Green's functions 
\begin{eqnarray}
&&D^R(t,t')=D^A(t',t) = -i \langle   P_{cl}(t) P_q (t')\rangle ~, \nonumber \\
&&D^K(t,t')=-i \langle   P_{cl}(t) P_{cl} (t')\rangle . 
\label{Green_defn}
\end{eqnarray}
where  $D^{R}$ and $D^{A}$ are the retarded and advanced response
functions of the oscillator mode, respectively, while $D^{K}$ is the
Keldysh Green's function, which contains information about the
temporal correlations and occupancy of the mode. The corresponding
Feynman diagrams for these Green's functions are  shown in Fig.\ref{green_vertex} where the classical and quantum fields are represented by solid and dashed lines, respectively.

 \begin{figure*}[t]
   \centering
  \includegraphics[width=0.99\textwidth]{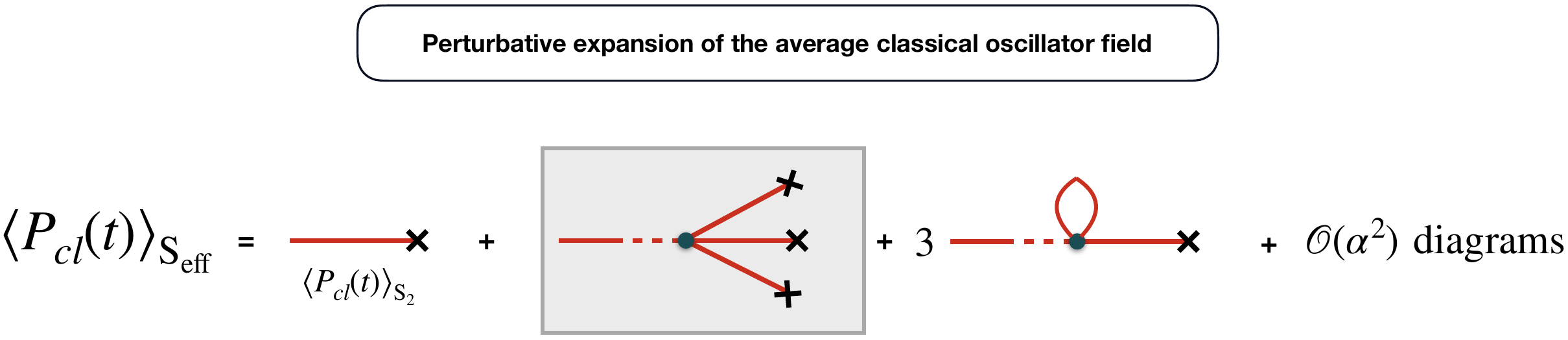}
  \caption{Diagrammatic expansion for the expectation 
of the classical component of the oscillator field:  obtained by expanding the quartic interaction term in the effective action ${\cal S}_{\rm{eff}}$ in powers of the coupling strength $\alpha$.  The first term in the R.H.S. (shown by a solid line ending at a cross) is the expectation value for the non-interacting oscillator given in Eq.\ref{P_cl_S2}. 
Among the first order diagrams, the shaded diagram contains maximum influence from the external electric field at this order of perturbation series while the un-shaded diagram contains loops of classical fields independent of $E(t)$. The former kind of diagrams form the ``tree series"  (see Fig.\ref{branch_EOM}) leading to classical EOM while the latter kind incorporates effects of quantum fluctuations in the dynamics that can not be captured within the classical theory.} 
   \label{diag}
   \end{figure*}

{\subsection{Saddle-point approach} \label{app:saddle}

In order to treat the Keldysh path-integral using 
saddle-point methods. we are required to vary the forward
and backwards time components of the Keldysh contour
independently. In the classical limit $\hbar  \rightarrow 0$, the action on
outward and return paths are extremized by the same classical path
\begin{equation}
\lim_{\hbar \rightarrow 0}
\langle P_{\pm } (t)\rangle  ={\cal P} (t),
\end{equation}
so that 
the outward and
return path actions are equal and the
Keldysh action on the classical contour is zero, ${\cal S}_{K}[{\cal P}]= ({\cal S}[P_{+}]-{\cal S}[P_{-}])\bigr|_{P_{\pm}={\cal P}}=0 $.
This means that a
variational approach must consider paths where $P_q\neq 0$, for which the Keldysh action is finite\cite{altlandbook}. 
  \begin{widetext}
The condition that the Keldysh action is stationary with respect to
independent variations of $P$ on the upper and lower contour
yields
\begin{eqnarray}
  -\frac{\delta {\cal S}_{K}}{\delta P(t)}
=  \left[\partial_{t}^{2} + \omega_P^2 + m(t)\right]P(t)+\alpha P^3(t)- E(t)=0,
    \label{wardidentity} 
\end{eqnarray}
{where $P(t)$  lies on  either the upper or lower part of the Keldysh contour.} This equation of motion defines the classical trajectory
${\cal P} (t)=
\langle  P_{cl} (t)\rangle_{\hbar \rightarrow 0}$ 
which is the saddle point of the Keldysh action.
Since the path integral $Z_{K}$ is invariant under a time-dependent
shift of variables, 
$P\rightarrow P + \delta P (t)$, which leaves the measure unchanged
${\cal D}[P]= {\cal D}[P+\delta P]$, 
\begin{eqnarray}
  0=  \int {\cal D}[P+\delta P] e^ {i {\cal S}_{K}[P+ \delta P]/\hbar}-\int {\cal D}[P] e^ {i {\cal S}_{K}[P]/\hbar}
  = 
 \int {\cal D}[P]e^{i {\cal S}_{K}[P]/\hbar} \int_{C} dt 
\left( 
\frac{i}{\hbar }
\frac{\delta {\cal S}_{K}}{\delta P (t)}\right) \delta P (t),
\end{eqnarray}
the equation of motion\eqref{wardidentity} 
is exact 
when averaged over quantum trajectories,
\begin{equation}
\left\langle \frac{\delta {\cal S}_{K}}{\delta P (t)}\right\rangle  =     \int {\cal D}[P]e^{i {\cal S}[P]/\hbar} \frac{\delta {\cal S}_{K}}{\delta P (t)}=0, 
\end{equation}
We should not be surprised, for 
\eqref{wardidentity} is equivalent to eliminating the momentum $\pi$ from the 
Heisenberg equations of  motion, ($ {\dot P} = -(i/\hbar)[H,P]=\pi \Rightarrow \ddot P = \dot \pi  = (i/\hbar)[H,\pi]=-\delta H/\delta P$). 
If we
 take the average of the upper and lower Keldysh contours, we obtain
\begin{equation}
\left(\partial_{t}^{2}+\omega_P^2+m(t)\right) \langle P_{cl}(t)
\rangle +\alpha \biggl\langle 
\frac{1}{2} ( P_{+}^{3}+ P_{-}^{3})
\biggr \rangle  =E(t),
    \label{cl_eom_again}
\end{equation}
where we assume a classical source field $E_{\pm } (t)=E (t)$.
We can expand
$ ( P_{+}^{3}+ P_{-}^{3})
 = 2 P_{cl}^{3}+
(3/2) P_{cl}P_{q}^{2}$ and by
rewriting 
the point-split
expectation value 
$\langle P_{cl} P_{q}^{2}\rangle \rightarrow \langle T_{C }\hat  P_{cl} (t_{1})\hat P_{q} (t_{2})\hat P_{q}
(t_{3})\rangle=0 $ in terms of time-ordered Heisenberg operators,
we find that it vanishes. At the Gaussian level of approximation, 
this can be understood because $\langle P^2_q P_{cl}\rangle = \langle P_q\rangle \langle P_q P_{cl}\rangle + \langle P_q^2\rangle \langle P_{cl}\rangle$, which vanishes because the first and second order moments of $P_q$ vanish, $\langle P_q\rangle= \langle P_q^2\rangle = 0$, but in Appendix \ref{AppC} we show that this is true to all orders.
It follows
that 
\begin{equation}
\left(\partial_{t}^{2}+\omega_P^2+m(t)\right) \langle P_{cl}(t) \rangle +\alpha \langle P_{cl} (t)^3 \rangle  =E(t).
    \label{cl_eom_again2}
\end{equation}
\end{widetext}

We  calculate the 
leading quantum fluctuations about the classical
trajectory $\langle  \delta P_{cl}^{2}\rangle \sim O (\hbar )$, determined
from the leading quadratic expansion of the 
action about the classical trajectory, 
\begin{equation} 
{\cal S}_{G}= 
\int_{C}dt \frac{1}{2}\left[\delta  \dot P^{2}-\biggl(\omega_{P}^{2}+m (t)+ 3 \alpha {\cal  P}^{2} (t)\biggr)\delta P^{2} \right]
\end{equation}
where $\delta P (t) = P (t) - {\cal  P} (t)$ 
is the deviation from the classical path. 
The cubic term in the equation of motion in Eq.\ref{wardidentity}
now acquires an additional component from the Wick contractions 
between the fluctuations, 
\begin{equation}
    \langle P^3_{cl}(t)\rangle_{{\cal S}_{G}}\rightarrow  {\cal P}(t)^3+  3  \langle  \delta P^2_{cl}(t)  
    \rangle_{{\cal S}_G} {\cal P}(t)\
\end{equation}
This introduces a self-energy  correction to the oscillator mass 
\begin{equation}\label{masren}
m(t) \rightarrow m(t) + \Sigma(t),
\end{equation}
where \begin{equation}
\Sigma (t)= 3 \alpha  \langle  \delta P^2_{cl}(t)  \rangle_{{\cal S}_G}=  3 i \alpha  D_{K} (t,t)
\end{equation}
is written in terms of the Keldysh Green's function $D_{K}
(t,t')= -i \langle \delta  P_{cl} (t)\delta P_{cl} (t')\rangle_{{\cal S}_{G}}$.
The self-energy correction to the mass modifies the equation of motion, 
   %
\begin{equation}
    \Ddot{\cal P}(t) + \left(\omega_P^2+m(t) + \Sigma (t)
 \right) {\cal P}(t) +\alpha {\cal P}(t)^3= E(t).
    \label{identity}
\end{equation}  
 Note that while the fluctuations are Gaussian, the classical
 equations of motion are nonlinear in $\alpha$. 
One of the key effects of this self-energy correction, is a 
shift in in the paraelectric to ferroelectric critical point.
}

\subsection{Keldysh Action}

We now re-interpret these results diagrammatically. 
The Keldysh action can be divided 
into Gaussian and quartic components, ${\cal S}_{K}={\cal S}_{2}+{\cal S}_{4}$. 
In the classical-quantum basis, 
\begin{eqnarray}
{\cal S}_2 &=& 
\frac{1}{2}\int_{-\infty }^{\infty}{dt}  \left (
{\begin{array}{cc}  P_{cl}(t), & P_q(t)  \end{array} }  \right) 
\Bigl[ {\begin{array}{cc}
  0 & D^{-1}_{A}\cr D^{-1}_{R} & D^{-1}_{K} 
  \end{array} }
\Bigr]
    \Big({\begin{array}{c} P_{cl}(t') \\ P_q(t')   \end{array} }  \Big)\cr
&+& \int  \limits_{t_0}^{\infty} dt  P_q(t) E(t),
    \label{S2_P}
  \end{eqnarray}
where the non-interacting inverse Green's functions are 
  \begin{equation}
D^{-1}_{R,A}=({{i}}\partial_t \pm{{i}}0^+)^2-(\omega_P^2+m(t)),
\end{equation}
while $D^{-1}_{K}$ 
is a purely imaginary term which sets the thermal boundary conditions.
The
quartic term 
\begin{eqnarray}
{\cal S}_4&=&-\frac{\alpha}{4} \int  \limits_{t_0}^{\infty} dt  \left[  P^4_+(t) - P^4_-(t)  \right] \nonumber \\
&=&-  \alpha \int  \limits_{t_0}^{\infty} dt \left [  P^3_{cl}(t)P_q(t)+ (1/4)P^3_{q}(t)P_{cl}(t)\right],
\end{eqnarray}
%
only  contains terms with odd powers of $P_{q}$, 
as shown in Fig.\ref{green_vertex}. 
This is 
the starting point for the diagrammatic expansions. 

 \begin{figure*}[t]
   \centering
  \includegraphics[width=0.99\textwidth]{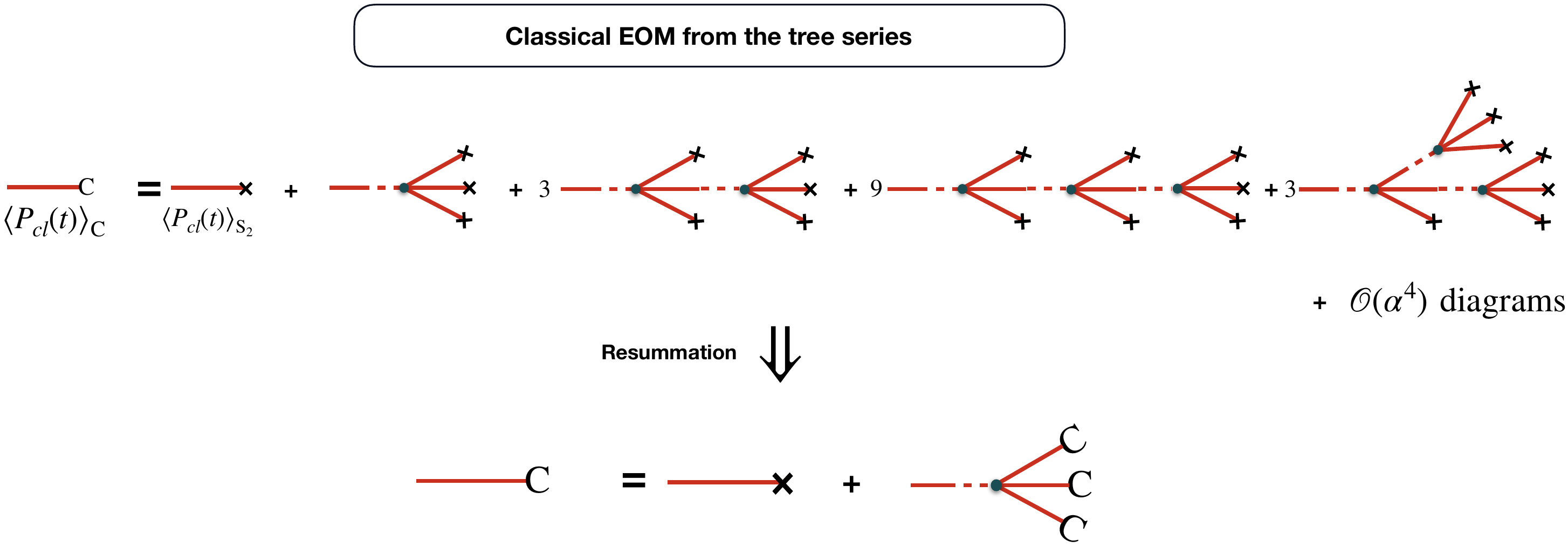}
  \caption{a)Recovery of the classical limit from Keldysh field theory:
  the solution of the classical EOM (see Eq.\ref{EOMFull}) $\langle
  P_{cl} \rangle_{\rm{C}} (t)$ is obtained from the full expectation
  value $\langle P_{cl} (t)\rangle _{\rm{S_{eff}}}(t)$ by restricting
  the sum only to the ``tree diagrams''   shown here. These are the diagrams which contain maximum powers of
  the external electric field, symbolically represented by a cross
  mark here, in each order of the perturbation theory. 
  These diagrams can be regrouped to yield the  non-perturbative classical solution.}
   \label{branch_EOM}
   \end{figure*}
The expectation value $\langle P_{cl} (t)\rangle _{\rm{S_{eff}}}$ can be expanded perturbatively in powers of the coupling strength $\alpha$ as
\begin{eqnarray}
  &&  \langle P_{cl} (t)\rangle_{{S_{K}}} =  \int D[P] ~e^{{{i}} ({\cal S}_2+{\cal S}_4)} P_{cl}(t) \cr
    &=& \langle P_{cl} (t)\rangle _{\rm{{\cal S}_{2}}} +   i\langle P_{cl} (t) {\cal S}_4  \rangle _{\rm{{\cal S}_{2}}} -   
     \frac{1}{2}\langle P_{cl} (t) \left({\cal S}_4 \right)^2 \rangle _{\rm{{\cal S}_{2}}} + ....,
    \label{P_cl_pert}\cr &&
\end{eqnarray}
where the expectation values are evaluated with respect to the
Gaussian action ${\cal S}_2$ \eqref{S2_P}. 
The Wick expansion of these terms involves the contraction of pairs of
$P$ fields into propagators and contractions of $P$ fields with the
external field $E$, giving rise to 
a series of Feynman
diagrams, as 
shown in Fig.\ref{diag}. The contraction of $P_{cl}$ with the external
field in the first term 
defines the linear response
\begin{equation}
     \langle P_{cl} (t)\rangle _{\rm{{\cal S}_{2}}} = - \int dt_1 D^R(t,t_1) E(t_1)
     \label{P_cl_S2},
\end{equation}
represented by a solid (classical) line ending at a
cross representing the electric field. Wick contractions of the  second term in
\eqref{P_cl_pert} generate two sets of diagrams: a ``tree diagram'',
involving three contractions of $P_{cl}$ with the external field, and
a ``Hartree diagram'', involving the contraction of two $P_{cl}$
fields. 

Next, we organize the higher order diagrams in Fig.\ref{diag} into two
classes: (a) ``tree diagrams" with a maximum number of classical
fields contracted with the external electric fields $E(t)$ and (b)
Hartree diagrams (unshaded in Fig.\ref{diag}) where a pair (or more)
of classical fields are contracted among themselves, forming a loop
which does not contain $E(t)$.  An example of the first class of
diagrams is shown by the gray shaded diagram in Fig.\ref{diag}. These
diagrams have maximum power of the electric field at a given order of
the perturbation series and contain only the interaction vertices with
$3$ classical fields and only $1$ quantum field. The second class of
diagrams involve the scattering off quantum fluctuations
. 
These scattering
processes 
describe the self-energy corrections to the 
mass of the soft polar mode by
quantum fluctuations, and can contain retarded and Keldysh Green's functions.

In Fig. \ref{branch_EOM}a), we show that a resummation of the
tree-diagrams leads to the classical EOM, whose solution is denoted
${\cal P}(t)$ (also given in
Eq. \ref{EOMFull}). 
To understand the resummation, we start with the
first order diagram in Fig. \ref{branch_EOM} (same as the gray shaded
diagram in Fig. \ref{diag}), where each of the three classical fields
of the interaction vertex are contracted with $E(t)$, yielding $
\left( \langle P_{cl} \rangle _{\rm{S_{2}}} \right)^3$. 
Higher order diagrams can be understood as the result of adding
further ``tree corrections'' to each external line. 
The resummation of these diagrams
non-perturbative classical solution can then be re-written in terms of Green's functions as,
\begin{eqnarray}
{\cal P} (t) = - \int  \limits_{t_0}^{\infty} dt_1 { D}^R(t,t_1) 
( E(t_1)- \alpha {\cal P} (t_{1}) ^3) .
\label{CL_eom}
\end{eqnarray}
This classical solution is represented diagrammatically by a solid line ending at the symbol $\rm{C}$ in Fig.\ref{branch_EOM}. This identification of the tree series as ``classical" diagrams is crucial to identify and study the quantum effects near the PE-FE transition which we discuss next.

\subsection{Perturbative Quantum Corrections } \label{Q_correction}

 In this section, we will study the leading quantum correction to the soft-mode mass, determining the resulting shift in the critical point, first for the case of single phonon mode (single nonlinear oscillator) in subsection \ref{singlemodeQ}, generalizing the calculation to the multi-mode case in subsection \ref{multimodeQ}.

\subsubsection{Single phonon mode} \label{singlemodeQ}

\begin{figure}[h] \includegraphics[width=0.95 \columnwidth]{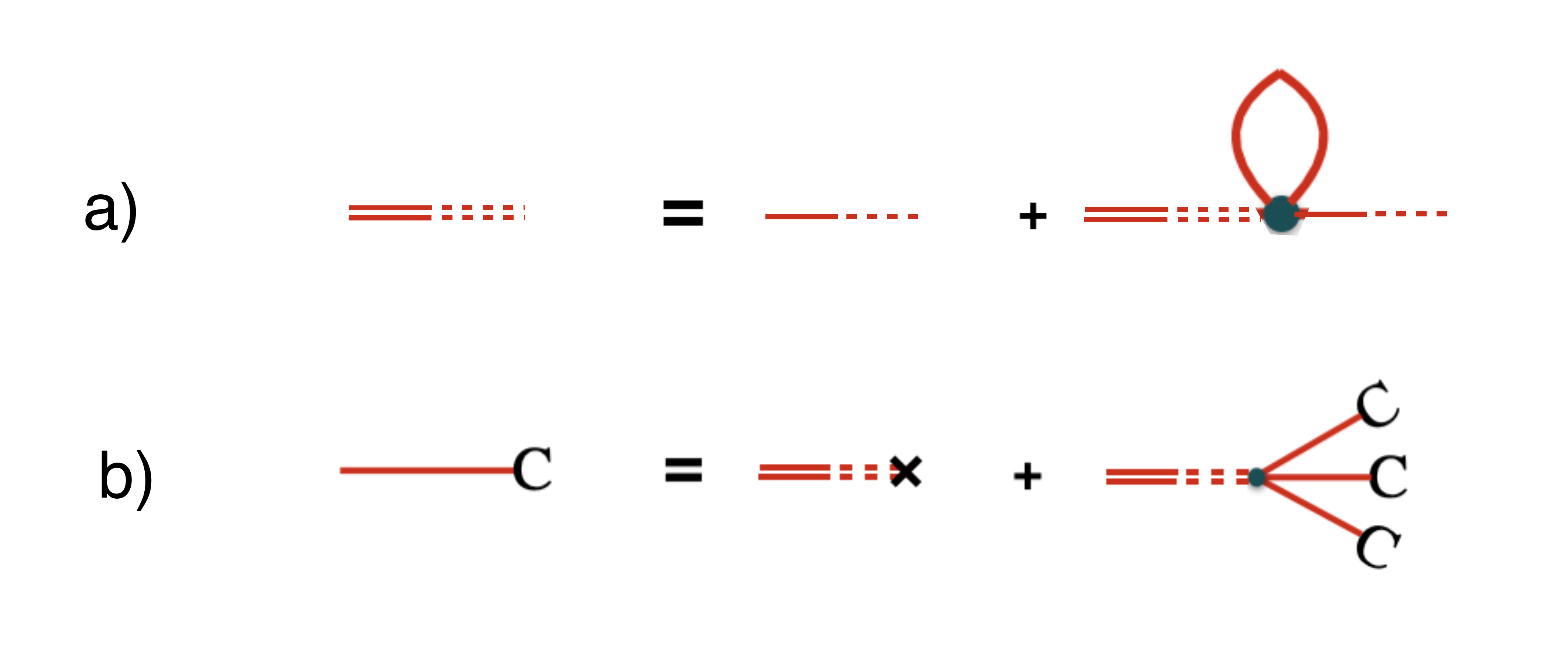}
\caption{a) Hartree self-energy insertion to the retarded propagator,
describing the leading $O (\hbar )$ effect of quantum fluctuations. 
b) The equation of motion now involves
the renormalized propagator.  
} \label{Q_corr}
\end{figure}

Quantum corrections to the classical
equations of motion are obtained by inserting self-energy corrections
to the retarded propagator. 
The leading Hartree self-energy  correction 
is derived from the one-loop retarded self-energy
$\Sigma^R(t)$,corresponding to the Hartree approximation,  as shown in Fig. \ref{Q_corr}. 
Within this approximation $\Sigma^R(t,t')= \Sigma (t)\delta(t-t')$ is 
local in time so that quantum fluctuations manifest themselves as a
time-dependent modification of the oscillator mass (see Appendix
\ref{twotime} for more details), 
\begin{equation}
\label{mass_onetime}
m(t) \rightarrow m(t) +  \Sigma(t)
\end{equation}
in Eq.\ref{remass}, where 
\begin{equation}
\Sigma (t)= 3 \alpha \langle  \delta P_{cl} (t)^{2}\rangle_{S_G} .
\end{equation}
{is proportional to the fluctuations in $P_{cl}(t)$ calculated using the Gaussian correction to the action, 
\begin{equation}
    {\cal S}_{G}= 
\int_{C}dt \frac{1}{2}\left[\delta  \dot P^{2}-\biggl(\omega_{P}^{2}+m (t)+ 3 \alpha {\cal  P}^{2} (t)\biggr)\delta P^{2} \right]
\end{equation}
However, if we restrict ourselves to the "para-electric phase" where
${\cal P}(t)= \langle P_{cl}(t)\rangle=0$, then in this case, the 
$3 \alpha {\cal  P}^{2} (t)$ term vanishes in ${\cal S}_G$. This does not restrict our consideration since in this section the main effect of the driving is incorporated in $m(t)$. The 
Gaussian action then coincides with the quadratic action ${\cal S}_G\equiv{\cal S}_2$ and the quantum corrections are the perturbative Hartree corrections, i.e
\begin{equation}
\Sigma (t)= 3 \alpha \langle  \delta P_{cl} (t)^{2}\rangle_{S_2}, \qquad ({\cal P}(t)=0).
\end{equation}
}


The quantum contribution to the self-energy is determined from the equal-time
Keldysh propagator so that 
\begin{equation}
    \Sigma(t) = {3} \alpha  i  D^K(t,t).
\end{equation}

Next, we calculate the equal time Keldysh Green's function $D^K(t,t)$
for the the non-interacting harmonic oscillator with a time-dependent
mass $\omega_P^2+m(t)$. 
To do so, we  rewrite the Keldysh Green's functions in terms of the
Heisenberg position operators of the Harmonic oscillator,
%
%
\begin{equation} 
D_{K} (t,t') = -i \langle  P_{cl} (t)P_{cl} (t')\rangle  =
-\frac{i}{2} \langle  \{ \hat P (t),\hat P (t') \}\rangle .
\end{equation}
Consider a non-interacting oscillator with time-dependent mass
$\omega^{2} (t)=\omega_P^2+m(t)$, where $m (0)=0$ and Hamiltonian 
\begin{equation} 
H = \frac{1}{2}\left (\hat \pi^{2}+ {\omega^{2}(t)}\hat P^{2}\right).
\end{equation}
Here $\hat P $ and $\hat \pi$ are canonical position and momentum
operators respectively. We now calculate $D_{K} (t,t')$ from the
expectation value of the Heisenberg operators $P (t)$, evaluated in
the initial state. We can relate the Heisenberg 
Schr\" odinger operators $P=P(0)$ and $\pi=\pi (0)$ by
\begin{equation} 
\hat P (t) = a(t)\hat P + b(t) \hat  \pi.
\end{equation}
From the equations of motion $\partial_{t}\hat P(t) = \hat \pi(t)$ and
$\partial_{t} \hat \pi(t) = - \omega^{2} (t)\hat  P(t)$, we deduce that $\partial^2_t P(t)= - \omega^2(t) P(t)$, so  
the coefficients $a(t)$ and $b(t)$ satisfy the differential equation%
\begin{eqnarray}
\left[\partial_{t}^{2}+ \omega_P^2+m(t) \right]
\begin{pmatrix}
a (t)\cr b (t)
\end{pmatrix} 
= 0,  
 \label{THO_DE}
\end{eqnarray}
subject to the boundary conditions  
\begin{equation} 
\pmat{a (0)\cr {b} (0)}= \pmat{1\cr 0}, \ 
\pmat{\dot{a} (0)\cr \dot{b} (0)}= \pmat{0\cr 1}, 
\end{equation}
The Keldysh Green's function for a system initially in the state 
$|n \rangle$ with $n$ phonons, can then be evaluated as 
\begin{eqnarray}\label{DK_THO}
D_{K} (t,t')&=&  -\frac{i}{2} \langle n|\{\hat{P}(t),\hat  P (t')\} |
n \rangle\cr & =& -i [a (t)a (t')\langle n \vert 
    P^{2}\vert n\rangle + b (t)b (t')\langle n \vert  \pi^{2}\vert n\rangle ]\cr
&=&  -i [a(t)a (t') + \omega_P^2 b(t)b (t')] \left(\frac{
n+\frac{1}{2}  
}{ \omega_P}\right).
\end{eqnarray}
so that 
\begin{equation} 
\langle  P_{cl}^{2} (t)\rangle  = (a^{2}(t) + \omega_P^2 b^{2}(t)) \left(\frac{
n+\frac{1}{2}  
}{ \omega_P}\right).
\end{equation}

As an example,  consider a linear time-dependence in the mass of the form $m(t)=-\omega_P^2 (t/t_0)$ for which the oscillator undergoes a quantum phase transition as $t \rightarrow t_0$ (at the bare level without self-energy corrections). We can obtain an analytical solution from Eq.\eqref{THO_DE} and in the extreme limits, this leads to simple form of the quantum correction (see Appendix
\ref{AppD} for more details) given by
\begin{eqnarray}
    \Sigma(t)=
\left\{
\begin{array}{lr}
  \frac{{ 3}\alpha}{{ 2}\sqrt{\omega_P^2+m(t)}} ,& \omega_P \gg
  1/t_0,  \cr
\frac{{ 3}\alpha}{ { 2}\omega_P} \left( 1+  \frac{t^3 \omega_P^2}{3t_0} \right),&\omega_P \ll 1/t_0,
\end{array}
 \right.
    \label{Signlemode_limit}
\end{eqnarray}
%
where the initial state of the oscillator is chosen to be the vacuum state $n=0$. In the next subsection, we will extend this analysis to the case of interacting phonons with different momenta and study experimentally measurable effects of the quantum fluctuations in the dynamics.

\subsubsection{Interacting phonons in 3D}\label{multimodeQ}

We now extend the discussion to the higher dimensional case where the
soft-phonon mode develops dispersion. As an illustration,  consider 3D
phonons with dispersion
\begin{equation}
\omega_k^2 (t)=\omega_P{^2}
+m(t)+(c k)^2
\label{eq:omk}
\end{equation}
with an ultraviolet momentum cutoff $\Lambda$ of the slow mode arising
from an underlying lattice. We further assume that the separation of
the energy scales between the slow $P$ modes and the fast $Q$ modes is
valid at all momenta, i.e. $c\Lambda \ll \Omega$. This allows us to
extend the effective potential approach for slow phonon modes
described in \eqref{EOMFull2} to the multi-mode ($P$
modes) case, where the resonantly driven $Q$ modes only result in a
time-dependent potential for $P$ modes modifying the bare mass
$\omega_P{^2}\rightarrow \omega_{P}^{2}+m (t)$. 


The quantum correction can be calculated from $\Sigma(t)={ \int_{\vec{k}} } \Sigma(\omega_k)$ where $\int_{\vec{k}}= \int d\vec{k}/(2\pi)^3$  and $\Sigma^R_q(\omega_k)$ is the retarded self-energy for the independent oscillator modes of frequency $\omega_k^2(t)$ \eqref{eq:omk},
\begin{equation}
    \Sigma(\omega_k) =  { 3} \alpha \frac{\left[a^2(k,t)+(\omega_P^2 + c^2k^2) b^2(k,t) \right]}{ \sqrt{\omega_P^2+c^2k^2}} \left(n+\frac{1}{2}  \right),
\end{equation}
where the coefficients $a(k,t)$ and $b(k,t)$ are calculated from Eq.\ref{THO_DE} replacing $\omega_P^2 +m(t)$ by $\omega_k^2(t)$. {Unless otherwise mentioned the initial state of the oscillators is chosen to be the vacuum state $n=0$.}

Before understanding the effects quantum fluctuations in the dynamics, we first focus on the equilibrium quantum correction $\Sigma \left(\sqrt{\omega_P^2+(c k)^2} \right)$ at $t=0$. This modifies the bare mass of the oscillators from $\omega_k^2 (t=0)$ to $\tilde{\omega}_k^2 (t=0)$ given by,
\begin{equation}
 \tilde{\omega}_k^2 (t=0)=\left[\omega_P^2+(c k)^2+{ \int_{\vec{k}}}\Sigma \left(\sqrt{\omega_P^2+(c k)^2} \right) \right] .
\end{equation}
Here, we note that ${\int_{\vec{k}} }\Sigma \left(\sqrt{\omega_P^2+(c k)^2} \right)$ is ultraviolet divergent ($ \propto \Lambda^2  \alpha /c$) and leads to a cut-off dependent shift of the zero-point energy of the oscillator. This divergence can be renormalized by a redefinition of the oscillator energy $\omega_P^2$ by $\tilde{\omega}_P^2= \omega_P^2+{  \int_{\vec{k}}  }\Sigma \left(\sqrt{\omega_P^2+(c k)^2} \right)\approx  \omega_P^2+ {3} \alpha \Lambda^2/(8\pi^2 c)  $. $\tilde{\omega_P}$ is the experimentally measurable energy of the phonon in equilibrium.

As time evolves, the dynamical quantum correction modifies the energy of the slow mode as,
\begin{eqnarray}
    \tilde{\omega}_k^2 (t)=\omega_P^2+(c k)^2+m(t)+{ \int_{\vec{k}} }\Sigma(\omega_k(t)). 
 \end{eqnarray}
To clearly differentiate between the $\Lambda$-dependence appearing in the equilibrium zero-point energy of the $P$ mode from the relevant dependence appearing in the dynamical quantum fluctuations, we rewrite the above equation as
\begin{eqnarray}    
 \tilde{\omega}_k^2 (t)   &=& \left[\tilde{\omega}_P^2+(c k)^2\right]
 +m(t) 
 \cr
    &+&{ \int_{\vec{k}} }\left[ \Sigma(\omega_k(t))-\Sigma^R_{q}\left(\sqrt{\omega_P^2+(c k)^2} \right)\right].
    \label{mass_t}
    \end{eqnarray}
Here, the 1st term of Eq. \ref{mass_t} corresponds to the dispersion of the oscillator at $(t=0)$ modified by the equilibrium quantum correction. 
The subsequent terms corresponds to the change in the oscillator energy at later time $t$ (compared to that at $t=0$). These terms consist of: (a) an explicit time-dependence through $m(t)$ induced by the external drive and (b) the change in the quantum self-energy (non-equilibrium quantum correction) $\delta\Sigma(t)={\int_{\vec{k}} }[\Sigma(\omega_k(t))- \Sigma(\omega_k(t=0))]$. By this rearrangement, we eliminate the  equilibrium ultraviolet divergences from the time-dependent part of the oscillator energy (2nd line of Eq. \ref{mass_t}). It is useful to recast Eq. \eqref{mass_t} in a form that does not include the unobservable bare phonon energy $\omega_P$, replacing the bare $\omega_P^2$ with the experimentally measurable $\tilde{\omega}_P^2$ in $\delta\Sigma(t) $ as,
\begin{eqnarray}
   && \tilde{\omega}_k^2 (t) \approx  \left[\tilde{\omega}_P^2+(c k)^2 \right]+m(t)  \nonumber \\
    &~&+{ \int_{\vec{k}} }\left[ \Sigma \left(\sqrt{\tilde{\omega}_P^2+(c k)^2+m(t)} \right)- \Sigma(\sqrt{\tilde{\omega}_P^2+(c k)^2})\right], \nonumber \\
    \label{mass_nocutoff}
\end{eqnarray}
which leads to corrections of higher order in $\alpha$. Indeed, replacing $\omega_P^2$ by $\tilde{\omega}_P^2$ leads to $\mathcal{O}(\alpha^2)$ change in $\delta\Sigma(t) $. As the leading order answer is $\mathcal{O}(\alpha)$, this effect can be neglected.

 \begin{figure}[t]
  \includegraphics[width=0.46\textwidth]{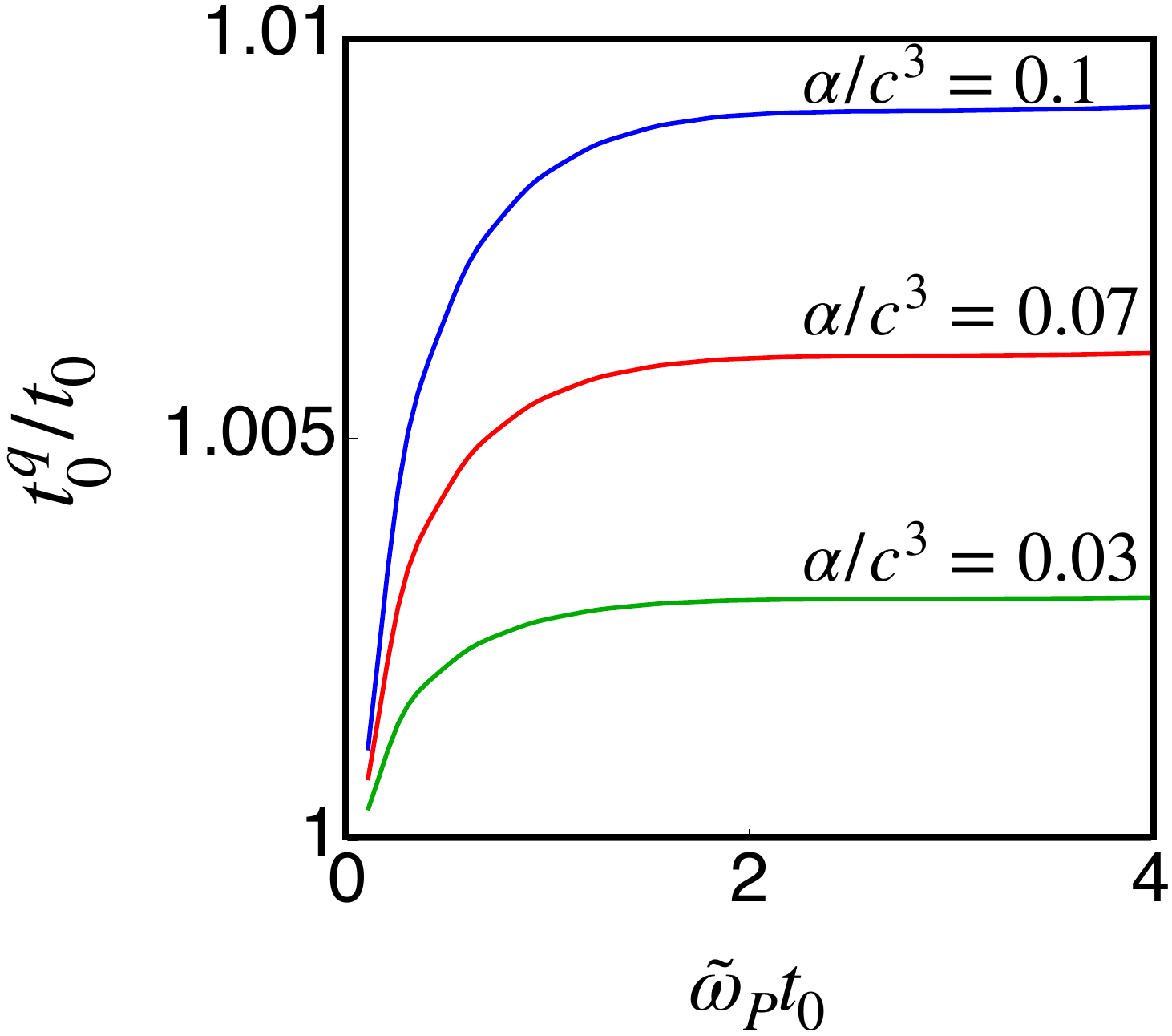}
  \caption{Shift in the critical point due to quantum fluctuations: the ratio between the critical time with quantum fluctuations and the classical critical time $t_0^q/t_0$ is plotted as a function of the rate of pumping expressed in the dimensionless unit $\tilde{\omega}_P t_0$ for three different strengths of the non-linearity (in dimensionless unit) $\alpha/c^3={0.03,0.07,0.1}$. In the fast pumping limit $\tilde{\omega}_P t_0 \ll 1$, $t_0^q$ grows from $t_0$ as $t_0^2$ with increasing $t_0$. This growth saturates in adiabatic limit $\tilde{\omega}_P t_0 \gg 1$  recovering the signatures of quantum criticality. { We choose $c^2\Lambda^2 / \tilde{\omega}_P^2 =100$ to set the ultraviolet cut-off $\Lambda$. } }
   \label{quantum_t0}
   \end{figure}

The non-equilibrium quantum correction leads us to predict  a shift in the critical fluence. To model this phenomenon, we model the effect of a time-dependent pump by $m(t)=-\tilde{\omega}_P^2 t/t_0$ corresponding to a linearly increasing fluence. In the absence of quantum corrections, the frequency would go to zero at $t=t_0$. To evaluate the effects of quantum fluctuations, we solve $\tilde{\omega}_k^2 (t_0^q)=0$ from Eq. \ref{mass_nocutoff} numerically to determine the shift in the critical time. In Fig.\ref{quantum_t0} we show the ratio of the shifted and bare critical times $t_0^q/t_0$  as a function of the rate of driving in the dimensionless unit $\tilde{\omega}_P t_0$ for three different values of the quartic coupling, $\alpha={0.03,0.07,0.1}$.

In the quantum quench limit, the system is driven to the QCP so rapidly that  system is unable adjust to the QCP ($1/\tilde{\omega}_P \gg t_0$ and $1/(c\Lambda) \gg t_0$); in this case
the dynamical quantum corrections are expected to be small. Using the analytical form given in Eq. \ref{Signlemode_limit} for the single mode case and generalizing it for the multi-mode case, we obtain an estimate of the quantum correction in the mass (in the limit $1/c\Lambda \gg t_0$ and $\tilde{\omega}_P \ll c\Lambda$),
\begin{equation}
    \delta\Sigma(t) \approx  {3}\frac{\alpha}{{ 16} \pi^2 c^3} \frac{t^3}{3t_0} c^4\Lambda^4.
\end{equation}
The above contribution from the non-equilibrium quantum correction is small for $t \sim t_0$ by $1/(c\Lambda) \gg t_0$. To find out the shift in the critical time, we solve
\begin{equation}
    \tilde{\omega}_P^2\left[ 1-\frac{t_0^q}{t_0} \right]+{ \delta }\Sigma (t_0^q) =0
    \label{Eq_tq}
\end{equation}
In the quantum quench limit, the leading order deviation of $t_0^q$ from $t_0$ is obtained from the above equation by replacing $\delta \Sigma (t_0^q)$ by $\delta \Sigma (t_0)$ to obtain,
\begin{eqnarray}
   \frac{t_0^{q}}{t_0} \approx 1+  
   \frac{\alpha}{16 \pi^2 c^3} \tilde{\omega}_P^2t_0^2 \left(  \frac{c^4\Lambda^4}{\tilde{\omega}_P^4} \right)
\end{eqnarray}
Thus the leading order deviation of $t_0^q$ from $t_0$ grows as $t_0^2$ as we increase the sweeping time $t_0$.

When the system is driven to the QCP slowly, the system has a longer time to adjust to the QCP and the effects of quantum corrections become marked (see Fig.\ref{quantum_t0}). 
In the adiabatic regime $\tilde{\omega}_Pt_0 \gg 1$, the quantum correction to the retarded self-energy takes the form,
\begin{equation}
    \Sigma(t) \approx  {3} \frac{\alpha }{{  8}\pi^2 c} \Lambda^2 \left[ 1-\frac{[\tilde{\omega}_P^2+m(t)]}{c^2 \Lambda^2}\log\left( \frac{c^2 \Lambda^2}{[\tilde{\omega}_P^2+m(t)]}\right) \right],
    \label{sigmaR}
\end{equation}
where we neglected the terms of $O(1)$ as small in comparison with the logarithm in the second term. As discussed above, the first term in \eqref{sigmaR} can be absorbed in a renormalization of the equilibrium parameters,
leading to an increase in the mode frequency due to 
\begin{widetext}
\begin{equation}
    \delta \Sigma (t)  = {3} \frac{\alpha }{{  8}\pi^2 c} \left[\frac{\tilde{\omega}_P^2}{2c^2 }\log\left( \frac{c^2 \Lambda^2}{\tilde{\omega}_P^2}\right) -\frac{[\tilde{\omega}_P^2+m(t)]}{2c^2 }\log\left( \frac{c^2 \Lambda^2}{[\tilde{\omega}_P^2+m(t)]}\right) \right].
\end{equation}
\end{widetext}
The resulting correction has a weak logarithmic dependence on $\Lambda$, consistent with a system at its the upper critical dimension, demonstrating that in the adiabatic limit we recover the signatures of equilibrium quantum criticality.

In the adiabatic limit, the leading order deviation of $t_0^q$ from $t_0$ is obtained from Eq. \ref{Eq_tq} by replacing $\delta \Sigma (t_0^q)$ by $\delta \Sigma (t_0)$ to obtain,
\begin{equation}
    t_0^q = t_0 \left[1  +  {3} \frac{\alpha\tilde{\omega}_P^2}{16 \pi^2c^3}\log\left( \frac{c^2 \Lambda^2}{\tilde{\omega}_P^2}
    \right) \right].
\end{equation}

We have computed an additional delay in the transition to the polar phase due to leading order quantum fluctuations.
Since here we are considering a model where the fluence varies linearly in time, this result corresponds
to an increase in the critical fluence.  More generally quantum fluctuations increase the renormalized mass, thus
requiring modified fluence profiles for the system to transition to the polar ordered state.
Therefore, the dependence of the critical fluence on the driving rate can be used to identify and to characterize quantum corrections in driven ferroelectrics.




\section{Summary}

In this work we have analyzed a model of a driven lattice system close to a ferroelectric instability. We have shown that classically, the driving can be described as a modification of the nonlinear phonon potential leading to a phase transition beyond a critical fluence. The structure of the ordered phase can be tuned by light polarization. For fluence above the critical one, a second phase transitions is possible that breaks additional symmetries. A further increase in fluence beyond a second critical value suppresses the ordered phase and in some cases, leads to a chaotic behavior. Beyond classical dynamics, we demonstrated that the classical equations of motion arise as a approximation to the full quantum Keldysh evolution and identified the lowest-order quantum corrections. The latter effects predict a dependence of the critical fluence on the driving rate, which may be observable experimentally. 


\section*{Acknowledgements}

We acknowledge stimulating discussions with G. Aeppli, A.V. Balatsky, J. Flick, W.Hu and M. Kulkarni. Z.Z. and P.Ch. are  
funded by DOE Basic Energy Sciences grant DE-SC0020353 and 
P. Co. is supported by  NSF grant DMR-1830707.  A. C. is a Rutgers Center for Materials Theory Abrahams Fellow
as was P.A.V. during his time at Rutgers when this project was initiated.

\appendix

\begin{widetext}

\section{Effects of Non-Biquadratic $P-Q$ Interactions}\label{app:PQeffects}

\begin{figure}
  \subfigure[]{
    \label{fig:ClPvsEApp1} 
    \includegraphics[height=2.5 in]{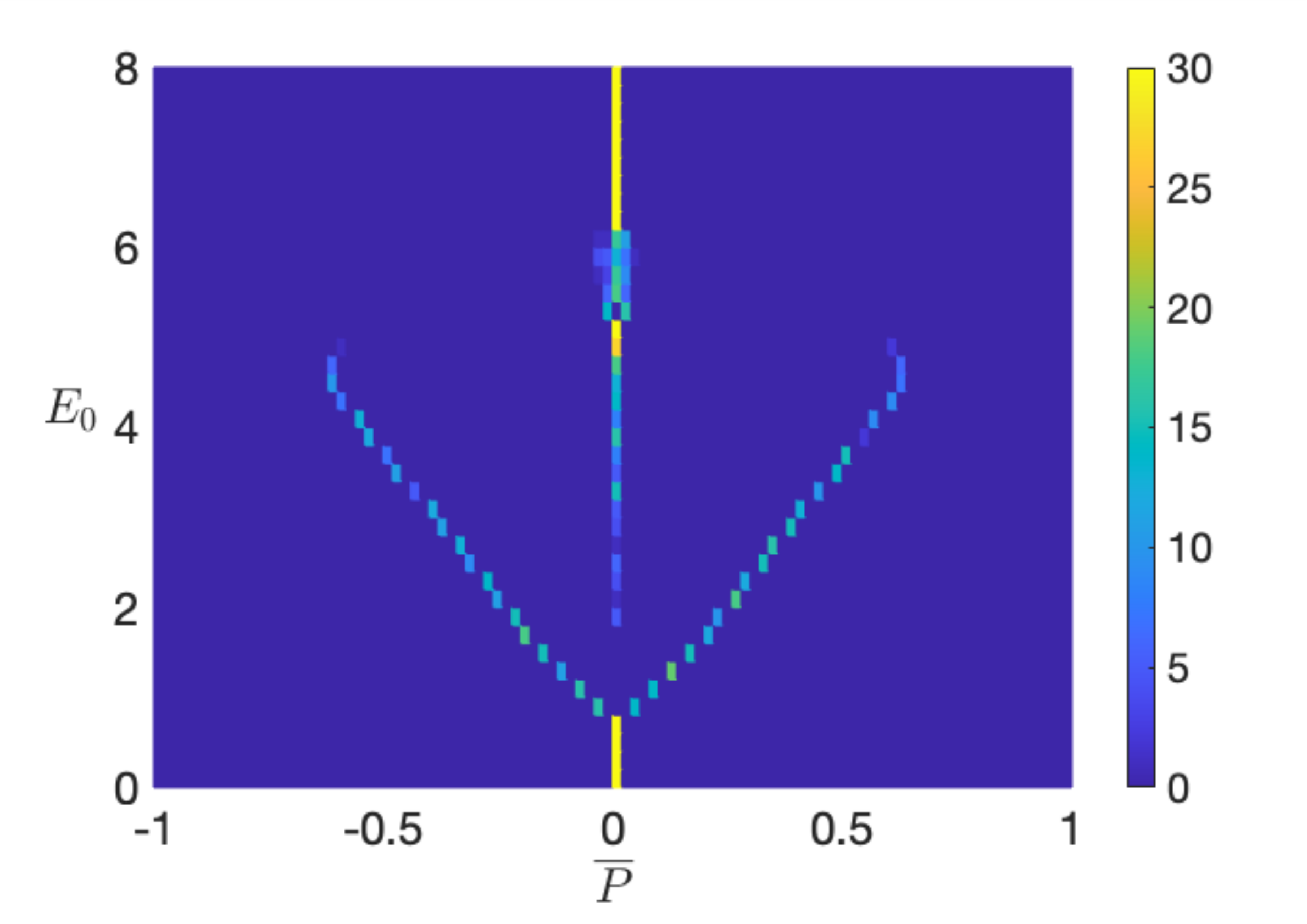}}
    \subfigure[]{
    \label{fig:CLPvsEApp2} 
    \includegraphics[height=2.5 in]{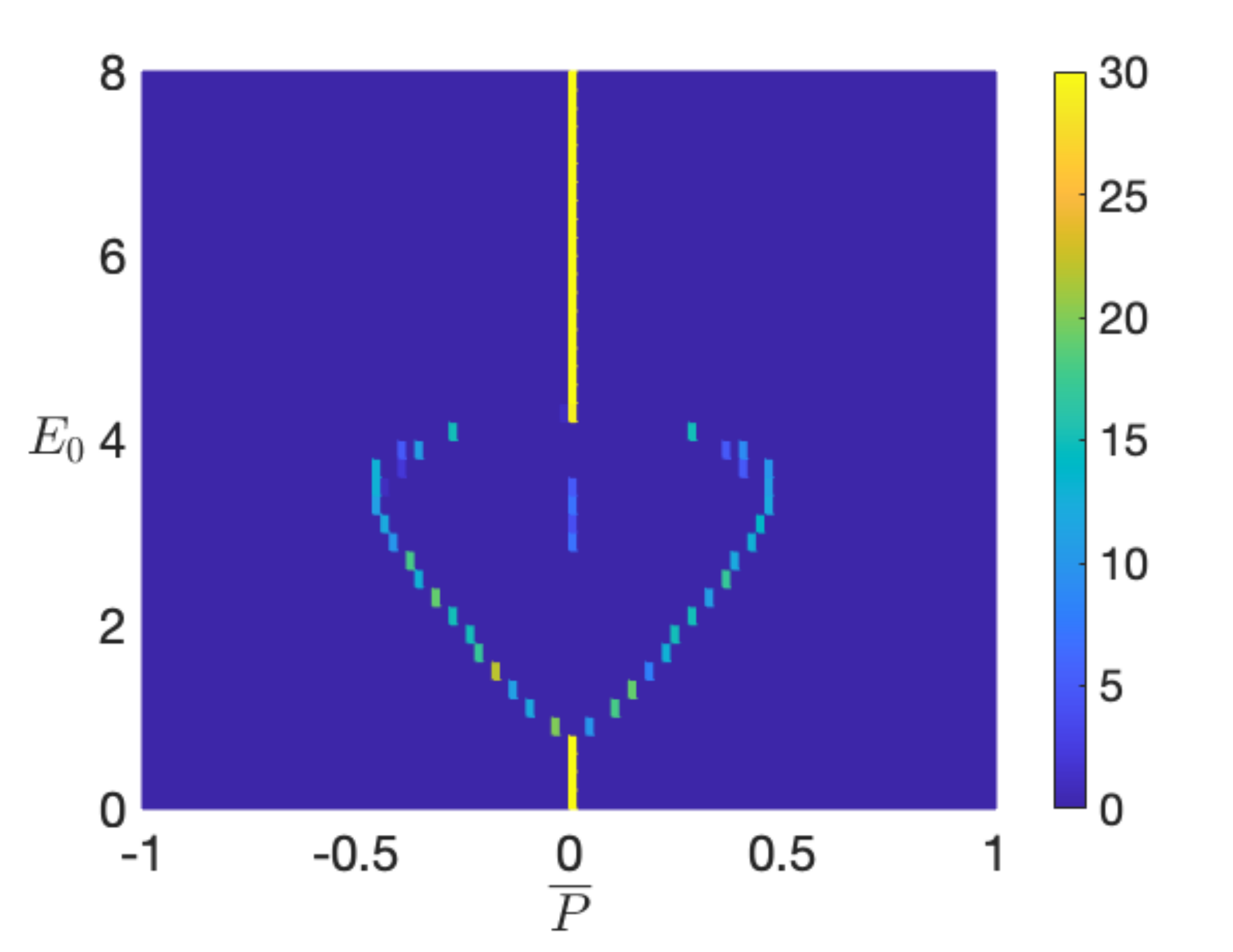}}
    \caption{
    Time-averaged polarization $\overline{P}$ versus electric field strength $E_0$ including non-biquadratic P-Q interactions [Eqs. (\ref{EOM:oddQ}) and (\ref{EOM:oddP})].
    These results were obtained from $30$ random initial conditions at each fixed $E_0$, and the color of grids indicates the number of times the system reaches $\overline{P}$ in its steady state.  The common parameter values for these plots were: $\omega_P=0.1, \omega_Q=2, \Omega=2.1, \gamma=0.01, Z_q=1, Z_p=0.3,$ $\alpha=1$ and $\beta_1=\beta_0=0.1$. We choose (a) $\gamma_{13}=\gamma_{31}=0.003$ and (b) $\gamma_{13}=\gamma_{31}=0.01$.}
\label{fig:ClassicalApp}
\end{figure}

In this appendix, we illustrate the reasons why non-biquadratic $P-Q$ interaction terms, such as $PQ$, $PQ^3$ and $P^3Q$, do not lead to qualitatively new effects and can be neglected. To start, we consider the two-oscillator model in Sec. \ref{TCOM} with additional action
\begin{equation}\label{Action:Sodd}
    S_\text{odd}=\int dt\left[ \gamma_{11}PQ+\frac{\gamma_{13}}{2}PQ^3+\frac{\gamma_{31}}{2}P^3Q\right].
\end{equation}
The classical EOM \eqref{EOM:P} then become

\begin{equation}
         \Ddot{Q}+\omega_Q^2 Q+\beta_1 \dot{Q}-\gamma P^2Q-\frac{3\gamma_{13}}{2}PQ^2-\frac{\gamma_{31}}{2}P^3-\gamma_{11}P=Z_q E(t),\label{EOM:oddQ}
\end{equation}    
\begin{equation}
       \Ddot{P}+\omega_P^2 P+\beta_0\dot{P}-\gamma Q^2P+\alpha P^3-\frac{\gamma_{13}}{2}Q^3-\frac{3\gamma_{31}}{2}P^2Q-\gamma_{11}Q=Z_p E(t). \label{EOM:oddP} 
    \end{equation}

\end{widetext}

If we first neglect the backaction terms in Eq. (\ref{EOM:oddQ}), namely the terms involving $P$s, then in this approximation we can apply (\ref{Qsteady}), namely
\begin{equation}
Q(t)=\chi_q E_0 \cos (\Omega t),
\end{equation}
where $\chi_q$ is defined in the main text. With this substitution,  the $Q$ term and $Q^3$ terms in Eq. (\ref{EOM:oddP})  can be directly absorbed into a time-dependent renormalization of the electric field,
\begin{multline}
     \tilde{E}(t)= E(t)\\+\frac{\gamma_{13}}{8Z_p}(\chi_q E_0)^3(\cos 3\Omega t+3\cos \Omega t)+\frac{\gamma_{11}}{Z_p}\chi_q E_0\cos (\Omega t).
\end{multline}
The additional two terms do not lead to any qualitatively new physics. The second term simply renormalizes the oscillating electric field at frequency $\Omega$, while the first one introduces oscillations at $3\Omega$ that are even further off-resonance for the $P$ mode. In addition, in the main text we have demonstrated that the critical $E_0$ for the light-induced transition scales with $\omega_P$ (see e.g. Eq. \eqref{eq:hessian_circ_1}) and is therefore small for a system close to the phase transition. In that regime, one can also justify the neglect of higher-order terms in $E_0$ due to the smallness of $E_0$. The $P^2Q$ term in (\ref{EOM:oddP})
corresponds to a time-dependent cubic potential for $P$ in the original action \eqref{Action:Sodd}. This term averages to zero over a period $2\pi/\Omega$ and is therefore unimportant in the paraelectric phase. In the ferroelectric phase, using the method we used to derive Eq. (\ref{ClSolution}), we can replace $P^2$ by $P_0^2$ allowing this term to also be absorbed into an effective electric field
\begin{equation}
    \tilde{\tilde{E}}(t)=\tilde{E}(t)+\chi_q E_0\frac{3\gamma_{31}}{2}P_0^2\cos (\Omega t).
\end{equation}
\\
We do not expect the omission of the backaction terms in Eq. (\ref{EOM:oddQ}) to change our arguments qualitatively, because the $P^2Q$ term merely results in a shift of the resonant frequency of the $Q$ mode, while the other backaction terms can be absorbed into an additional renormalization of the electric field, using similar arguments to those presented above.

To verify these arguments directly, Figure \ref{fig:ClPvsEApp1} and \ref{fig:CLPvsEApp2} show the numerical solutions of Eqs. (\ref{EOM:oddQ}) and (\ref{EOM:oddP}). Comparing them to Fig. \ref{fig:CLPvsEMesh2}, we see that the inclusion of non-biquadratic terms hardly affects the onset of ferroelectricity, although it can lead to quantitative differences at higher fields. However, we note that the qualitative features, such as multiple coexisting solutions and chaotic behavior, still persist when these terms are included.

\section{Two-Time Correlations in Quantum Corrections} \label{twotime}

\begin{figure}[h]
  \includegraphics[width=0.48\textwidth]{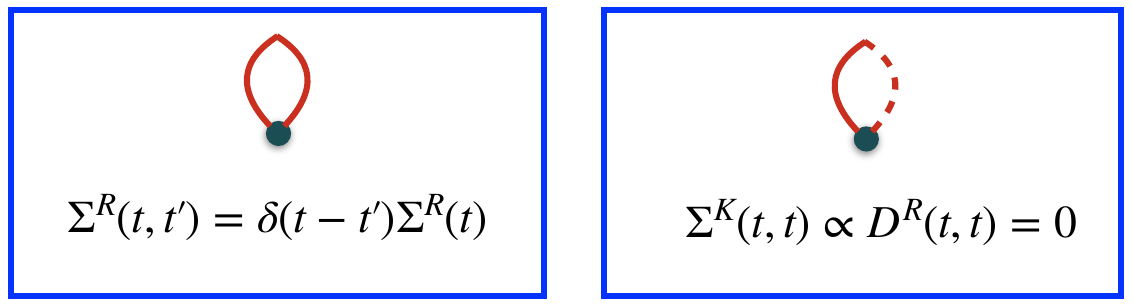}
  {\caption{Diagrams for retarded and Keldysh self-energy within Hartree approximation are shown. At the one-loop order, the retarded self-energy is a frequency independent one-time object $\Sigma^R(t)$ while the Keldysh component vanishes due to the causality structure of the equal time Green's functions. }} 
   \label{quantum_appendix}
   \end{figure}

In the main text, we have perturbatively calculated the quantum corrections to the classical EOM $\langle P_{cl} \rangle_C(t)$ of the oscillator given in Eq. \ref{CL_eom} by modifying the mass of the polar mode $m(t)$ in Eq.\ref{mass_onetime}. In this appendix, we present the details of calculations leading to this result. In particular, we show that in the lowest order (the one which we consider),  two-time correlations and noise terms do not arise from quantum corrections.


 The quantum correction is obtained via the retarded self-energy $\Sigma^R$ which can in general is a function of two times, $\Sigma^R(t,t')$ \cite{Grandi2021,Klein2020}. However, if we calculate $\Sigma^R(t,t')$ within a Hartree appproximation, i.e. restricting ourselves to lowest order diagrams (Fig. \ref{quantum_appendix}), the retarded self-energy becomes local in time,  $\Sigma(t,t')=\Sigma^R(t) \delta(t-t')$\cite{kamenevbook}. 
 The Keldysh self-energy $\Sigma^K(t,t')$ (Fig. \ref{quantum_appendix}, right) which typically introdues noise in the dynamics (see section 11.3 of the Ref.\onlinecite{altlandbook}),
 vanishes within a Hartree aprroximation due to causality structure of the equal-time Green's functions $D^R(t,t)=0$. Thus, within the Hartree approximation, we may write the non-equilibrium Dyson equation for the
interacting retarded Green's function as \cite{kamenevbook}
\begin{equation}
    D^R_{int}(t,t')= D^R(t,t') + \int  D^R(t,t_1) \Sigma^R(t_1) D^R_{int}(t_1,t').
\end{equation}
 Inverting the above equation by applying $D^{-1R}$ from the left and $D_{int}^{-1R}$ from the right we obtain,
 \begin{eqnarray}
   && D_{int}^{-1R}(t,t')=D^{-1R}(t,t') -\delta(t-t') \Sigma^R(t')\nonumber \\
    &&=\delta(t-t') \left[-\partial_t^2 - (\omega^2_P+m(t)+\Sigma^R(t)) \right]
\end{eqnarray}
In this way, the quantum corrections in the Hartree approximation appear as a modification in the time-dependence of the mass of the polar mode.

\begin{widetext}

\section{Proof that $\langle P_q^2 P_{cl}\rangle = 0$.}\label{AppC}

To prove the identity $\langle P_q^2 P_{cl}\rangle = 0$ we consider the point-split relation $\langle P_q(t+\delta)P_q(t)P_{cl}(t-\delta)\rangle$, which we rewrite as a path-ordered expectation value of the corresponding Heisenberg operators
\begin{eqnarray}
    \langle P_q(t+\delta)P_q(t)P_{cl}(t-\delta)\rangle = \frac{1}{2}\langle 
    {\cal T}_C(\hat P_+(t+\delta) - \hat P_-(t+\delta))  (\hat P_+(t) - \hat P_-(t)) (\hat P_+(t-\delta) + \hat P_{-}(t-\delta) )\rangle.\cr
\end{eqnarray}
where ${\cal T}_C$ denotes ordering along the Keldysh contour.
We now expand this into eight terms, noting that (i) operators on the lower (-) contour occur after operators on the upper contour (+), (ii) operators on the upper contour are time-ordered (iii) operators on the lower contour are reverse-time ordered.  Thus since $t+\delta > t > t-\delta$,
\begin{eqnarray}
 \langle   {\cal T}_C  
 \hat P_+(t+\delta)\hat P_+(t)\hat P_+(t-\delta)
 \rangle &=
 \phantom{-}
 & \langle  \hat P(t+\delta)\hat P(t)\hat P(t-\delta)\rangle 
 \cr
  \langle   {\cal T}_C  \hat P_+(t+\delta)\hat P_+(t)\hat P_-(t-\delta)\rangle &=\phantom{-}& \langle \hat P(t-\delta)\hat P(t+\delta)\hat P(t)   \rangle  
  \cr 
    -\langle{\cal T}_C \hat P_+(t+\delta)\hat P_-(t)\hat P_+(t-\delta)\rangle &=-& \langle \hat P(t)\hat P(t+\delta)\hat P(t-\delta) \rangle \cr 
      -\langle   {\cal T}_C  \hat P_+(t+\delta)\hat P_-(t)\hat P_-(t-\delta)\rangle &=-&\langle \hat P(t-\delta)\hat P(t)\hat P(t+\delta) \rangle \cr
        -\langle   {\cal T}_C  \hat P_-(t+\delta)\hat P_+(t)\hat P_+(t-\delta)\rangle &= -&\langle \hat P(t+\delta)\hat P(t)\hat P(t-\delta) \rangle \cr
          -\langle   {\cal T}_C  \hat P_-(t+\delta)\hat P_+(t)\hat P_-(t-\delta)\rangle &= -&\langle \hat P(t-\delta)\hat P(t+\delta)\hat P(t) \rangle \cr
            \langle   {\cal T}_C  \hat P_-(t+\delta)\hat P_-(t)\hat P_+(t-\delta)\rangle &=\phantom{-}& \langle \hat P(t)\hat P(t+\delta)\hat P(t-\delta) \rangle \cr
                \langle   {\cal T}_C  \hat P_-(t+\delta)\hat P_-(t)\hat P_-(t-\delta)\rangle &=\phantom{-}& \langle  \hat P(t-\delta)\hat P(t)\hat P(t+\delta)\rangle. 
\end{eqnarray}
We see that the first and fifth, second and sixth, third and seventh and 
fourth and eighth terms cancel one-another, so that the total sums to zero. 

\section{Derivation of Quantum Self-energy for single mode phonon}\label{AppD}
In the main text, we calculated the quantum correction in the form of Hartree self-energy $\Sigma(t)$ in Eq.\ref{Signlemode_limit} for a single-mode phonon. Here we will derive $\sigma(t)$ starting from the formulation to calculate Keldysh Green's function of a harmonic oscilator with time-dependent frequency given in Eq.\ref{DK_THO}. The Hartree self-energy,
\begin{eqnarray}
     \Sigma(t) & = & {3} \alpha  i  D^K(t,t) 
     =\frac{
3 \alpha  
}{2 \omega_P} (a^{2}(t) + \omega_P^2 b^{2}(t)) 
\end{eqnarray}
We solve the differential equations given in Eq.\ref{THO_DE} to obatin the solutions with $z=\omega_P t_0$,
\begin{eqnarray}
    a^{2}(t)&=&\pi^2 \left[Ai'(-z^{2/3}) Bi\left\{ \left(1-\frac{t}{t_0}\right)z^{2/3} \right\}-  Bi'(-z^{2/3}) Ai\left\{- \left(1-\frac{t}{t_0}\right)z^{2/3} \right\} \right]^2 \nonumber \\
    b^{2}(t)&=& \pi^2 z^{2/3} \left[Ai(-z^{2/3}) Bi \left\{ -\left(1-\frac{t}{t_0}\right)z^{2/3} \right\}-  Bi(-z^{2/3}) Ai \left\{- \left(1-\frac{t}{t_0}\right)z^{2/3} \right \} \right]^2 
\end{eqnarray}
where $Ai$ and $Bi$ are Airy functions and $Ai'$ and $Bi'$ denotes their derivatives respectively.

In the quench limit, $\omega_P t_0 \ll 1$, the Airy functions can be expanded in power series of $z$ of the form,
\begin{eqnarray}
Ai(z) &&\approx \frac{1}{3^{2/3} \Gamma(\frac{2}{3})}\left(1+\frac{z^3}{6}\right) -  \frac{z}{3^{1/3}\Gamma(\frac{1}{3})}, \nonumber \\
Bi(z) &&\approx \frac{1}{3^{1/6} \Gamma(\frac{2}{3})}\left(1+\frac{z^3}{6}\right) -  \frac{z 3^{1/6}}{\Gamma(\frac{1}{3})}, \nonumber \\
Ai'(z) &&\approx \frac{1}{3^{1/3} \Gamma(\frac{1}{3})}\left(1+\frac{z^3}{3}\right) -  \frac{z^2}{2\times 3^{2/3}\Gamma(\frac{2}{3})}, \nonumber \\
Bi'(z) &&\approx \frac{3^{1/6}}{ \Gamma(\frac{1}{3})}\left(1+\frac{z^3}{3}\right) -  \frac{z^2 }{2\times 3^{1/6}\Gamma(\frac{2}{3})}, \nonumber \\
\end{eqnarray}
Collecting the coefficients of the leading powers of $z$ we obtain Eq.\ref{Signlemode_limit} as,
\begin{equation}
   \Sigma(t)= \frac{3\alpha}{ 2\omega_P} \left( 1+  \frac{t^3 \omega_P^2}{3t_0} \right),\omega_P \ll 1/t_0.
\end{equation}
On the other hand, in the adiabatic limit $\omega_P t_0 \gg 1$, we get the same answer as the time-independent harmonic oscillator with its mass term $\omega^2_P$ replaced by $\omega^2_P + m(t)$.
\end{widetext}

\bibliographystyle{apsrev4-2}

\bibliography{Reference}

\end{document}